\documentclass[aps,prl,letterpaper,superscriptaddress,twocolumn,amsmath,amssymb,longbibliography]{revtex4-2}
\usepackage[T1]{fontenc}
\usepackage[utf8]{inputenc}
\usepackage{float}
\usepackage{amsmath,amssymb,bm}
\usepackage{graphicx}
\usepackage[unicode=true,pdfusetitle,bookmarks=true,bookmarksnumbered=false,bookmarksopen=false, breaklinks=false,pdfborder={0 0 1},backref=false,colorlinks=false] {hyperref}
\usepackage{dsfont}
\usepackage{csquotes}
\usepackage{xcolor}

\newcommand{\eps}{\epsilon}
\newcommand{\Tr}{\operatorname{Tr}}

\newcommand{\Om}{\Omega}
\newcommand{\dg}{\dagger}

\newcommand{\Hqp}{H^{\mathrm{qp}}}
\newcommand{\Hemb}{\hat{H}^{\mathrm{emb}}}

\newcommand{\nF}{n_F}



\usepackage{babel}
\begin{document}
\title{Ghost rotationally invariant slave-boson approach to response functions}

\author{Tsung-Han Lee}
\affiliation{Department of Physics, National Chung Cheng University, Chiayi 62102, Taiwan}
\affiliation{Physics Division, National Center for Theoretical Sciences, Taipei 10617, Taiwan}

\author{Corey Melnick}
\affiliation{Physics and Astronomy Department, Rutgers University, Piscataway, New Jersey 08854, USA}
\affiliation{Condensed Matter Physics and Materials Science Department, Brookhaven National Laboratory, Upton, New York 11973, USA}

\author{Nicola Lanat\`a}
\affiliation{School of Physics and Astronomy, Rochester Institute of Technology, 84 Lomb Memorial Drive, Rochester, New York 14623, USA}
\affiliation{Center for Computational Quantum Physics, Flatiron Institute, New York, New York 10010, USA}

\author{Gabriel Kotliar}
\affiliation{Physics and Astronomy Department, Rutgers University, Piscataway, New Jersey 08854, USA}
\affiliation{Condensed Matter Physics and Materials Science Department, Brookhaven National Laboratory, Upton, New York 11973, USA}

\begin{abstract}
We propose a finite-temperature Gaussian-fluctuation formalism for response functions within ghost rotationally invariant slave-boson (gRISB) theory and apply it to the two-dimensional Hubbard model.
The gRISB magnetic phase boundary and static spin and charge susceptibilities capture the trends of dynamical mean-field theory (DMFT), significantly improving upon standard RISB and the random phase approximation (RPA).
At the RISB level, our formalism recovers the Kotliar--Ruckenstein response.
The dynamical susceptibilities resolve paramagnons, low-energy sound modes, and high-energy Hubbard excitations.
Because they are evaluated directly on the real-frequency axis, the results avoid Matsubara truncation, stochastic sampling, and numerical analytic continuation.
We also introduce a conserving quasiparticle approximation that expresses the response in terms of dressed Green's functions and vertex corrections.
\end{abstract}
\maketitle

\emph{Introduction}.--- Computing response functions of strongly correlated electron systems remains an outstanding challenge in condensed matter physics, while their real-frequency structure is directly probed by inelastic neutron and resonant inelastic X-ray scattering.
The random phase approximation (RPA) is inexpensive but overestimates spin fluctuations and becomes unreliable at strong coupling~\cite{Moriya1985,Schafer2021}; extensions such as the fluctuation-exchange approximation reach moderately correlated systems~\cite{Bickers1989,Vilk1997}.
Dynamical mean-field theory (DMFT) and its diagrammatic extensions provide a nonperturbative route for model Hamiltonians and real materials~\cite{DMFT_RMP_1996,DMFT_RMP_2006,Rohringer2018}, but computing two-particle quantities requires measuring and inverting two-particle correlation functions and, depending on the impurity solver, can be hindered by statistical noise or a fermionic sign problem~\cite{Gull2011}.
Efficient complementary approaches are therefore needed.

The rotationally invariant slave-boson (RISB) approach and the equivalent Gutzwiller approximation offer an efficient route to strongly correlated systems~\cite{Lecherman_2007,Bunemann_2007,Lanata_2015_PRX,Lanata_2017_PRL} and have been extended to response functions~\cite{Seibold2001_charge,Seibold2004_magnetic,von_Oelsen_2011_PRL,Lee_PRX_2021}.
Its single-orbital variant, the Kotliar--Ruckenstein slave-boson (KRSB) mean field, computes responses from Gaussian fluctuations~\cite{Kotliar1986,Lavagna1990,Li_1989_PRB_SRISB,Li1991,Zimmermann1997} and already captures high-energy Hubbard excitations in the charge channel~\cite{Dao2017}.
Nevertheless, the accuracy of RISB for strongly correlated response functions remains insufficiently established.
The ghost extension (gRISB)~\cite{gRISB_2017,gRISB_2021,gRISB_2022} systematically improves the single-particle accuracy toward DMFT~\cite{Lee_gRISB_2023_1,Lee_gRISB_2023_2}, has recently been formulated at finite temperature and recast as a DMFT Baym--Kadanoff functional with a pole-expansion self-energy~\cite{giuli2026}, and has been applied to multiorbital models and real materials~\cite{Carlos_multiorb_2023,Lee_gRISB_2023_2,Lee_gRISB_2024}.
These developments motivate extending gRISB to two-particle response functions.

In this work, we propose a dynamical extension of the finite-temperature gRISB functional of Ref.~\cite{giuli2026}, in which the variational fields are promoted to space- and imaginary-time-dependent paths, and evaluate the Gaussian response functions around its saddle point~\cite{supplement}.
For the two-dimensional Hubbard model, the method reproduces the DMFT trends in the magnetic phase boundary and static spin and charge susceptibilities, improving markedly upon RISB and RPA.
At the RISB level, it recovers the KRSB response, thereby bridging the KRSB and DMFT descriptions.
The dynamical susceptibilities resolve paramagnons, low-energy sound modes, and high-energy Hubbard excitations directly on the real-frequency axis.
A conserving quasiparticle approximation further provides a diagrammatic representation in terms of dressed Green's functions and fluctuation-generated vertices.

\emph{Formalism.}--- We consider the one-orbital Hubbard model:
\begin{align}
H & =\sum_{\mathbf{k}\sigma}\epsilon_{\mathbf{k}}c_{\mathbf{k}\sigma}^{\dagger}c_{\mathbf{k}\sigma}+\sum_{i}\big[Uc_{i\uparrow}^{\dagger}c_{i\uparrow}c_{i\downarrow}^{\dagger}c_{i\downarrow}-\mu\!\sum_\sigma c_{i\sigma}^{\dagger}c_{i\sigma}\big],\label{eq:H}
\end{align}
where $\epsilon_{\mathbf{k}}=-2t(\cos k_{x}+\cos k_{y})$ is the nearest-neighbor dispersion of the square lattice with hopping amplitude $t$, and $\mu$ is the chemical potential. Here $\sigma$ is the spin index, $\mathbf{k}$ is the momentum label, and $i$ is the site index. Throughout, we set $t=1$ as the energy unit.

Our starting point is the established static finite-temperature gRISB functional of Ref.~\cite{giuli2026}. We propose to promote its fields to periodic imaginary-time paths, which gives the dynamical Lagrangian
\begin{equation}
\mathcal{L}[\bm x(\tau)]
 = \mathcal{L}_{\mathrm{qp}}[\bm x(\tau)]
 + \mathcal{L}_{\mathrm{emb}}[\bm x(\tau)]
 - \mathcal{L}_{\mathrm{mix}}[\bm x(\tau)]
 \,,
 \label{eq:dyn-lagrange}
\end{equation}
where the independent components of the variational matrices
are collected in
\begin{equation}
\bm x_i(\tau)
=
\big(R_i',R_i'',\lambda_i,D_i',D_i'',\lambda_i^c\big)(\tau).
\label{eq:xvec}
\end{equation}
Here $i$ labels lattice sites, $\tau\in[0,\beta)$ is imaginary
time, $R_i=R_i'+iR_i''$, and
$\mathcal D_i=D_i'+iD_i''$.
For static fields, Eq.~\eqref{eq:dyn-lagrange}
reduces to the equilibrium functional of Ref.~\cite{giuli2026}, and the response follows from its
Gaussian expansion about the saddle point~\cite{supplement}. A rigorous derivation of Eq.~\eqref{eq:dyn-lagrange} may be sought from
the time-dependent Dirac--Frenkel variational
principle~\cite{Schiro2010,Fabrizio2013,Fabrizio2017,Guerci2023}, from a slave-boson path-integral
formulation~\cite{Arrigoni1995,Riegler2026}, or from the DMFT response formalism with a discretized
bath, in the spirit of Ref.~\cite{giuli2026}.

The first term is:
\begin{equation}
\mathcal{L}_{\mathrm{qp}} = -\frac{1}{\beta N}
  \ln\Tr\,\mathcal{T}_\tau\exp\!\Big[-\!\int_0^{\beta}\! d\tau\,
  \hat H^{\mathrm{qp}}(\tau)\Big],
\label{eq:Lqp}
\end{equation}
where
\begin{align}
\hat H^{\mathrm{qp}}(\tau)&=\underset{\bm k_1\bm k_2 ab}{\sum}
 \big[\underline H^{\mathrm{qp}}(\tau)\big]^{ab}_{\bm k_1\bm k_2}\,
 f^\dg_{\bm k_1 a} f_{\bm k_2 b},
 \\
\big[\underline H^{\mathrm{qp}}(\tau)\big]^{ab}_{\bm k_1\bm k_2}
 &= \sum_{\bm k_3 \sigma}\big[R_{\bm k_1-\bm k_3}(\tau)\big]^{a\sigma}
    \big[\eps_{\bm k_3}\big]^{\sigma}
    \big[R^\dg_{\bm k_3-\bm k_2}(\tau)\big]^{\sigma b} \nonumber\\
 &\quad + \big[\lambda_{\bm k_1-\bm k_2}(\tau)\big]^{ab}
\label{eq:Hqpmat}
\end{align}
is the so-called quasiparticle Hamiltonian, $f_{\bm k a}$ are the auxiliary fermionic modes with orbital index $a=1,\ldots,N_a$
and momentum $\bm k$, $[\eps_{\bm k}]^{\sigma}\equiv\eps_{\bm k}$ the spin-diagonal dispersion, $R$ the renormalization matrix, $\lambda$ the quasiparticle potential, $\beta$ the inverse temperature, $N$ the number of lattice sites, and $\Tr$ the many-body Fock-space trace.
The number of auxiliary orbitals controls the accuracy of the method, which reduces to the standard RISB for $N_a=2$. Throughout this paper we set $N_a=6$ for gRISB.

The second term is:
\begin{equation}
\mathcal{L}_{\mathrm{emb}} = -\frac{1}{\beta N}\sum_i
  \ln\Tr\,\mathcal{T}_\tau\exp\!\Big[-\!\int_0^{\beta}\! d\tau\,\Hemb_i(\tau)\Big]
  \,,
\end{equation}
where
\begin{align}
\Hemb_i(\tau) &= \hat{H}_{\mathrm{loc}}[c^\dg_i, c_i]
 + \sum_{a\sigma}\!\big(\mathcal{D}_{i,a\sigma}(\tau)\, c^\dg_{i\sigma} b_{ia}
   + \mathrm{H.c.}\big) \nonumber\\
 &\quad + \sum_{ab}\lambda^c_{i,ab}(\tau)\, b_{ib} b^\dg_{ia}
 \,,
\label{eq:Lemb}
\end{align}
is the so-called interacting embedding Hamiltonian,
where $\hat{H}_{\mathrm{loc}}$ contains the local interaction and chemical-potential terms, $\mathcal{D}$ is the impurity--bath hybridization, $\lambda^c$ is the bath potential, and $b_{ia}$ are the auxiliary bath fermions.

The last term is:
\begin{equation}
\mathcal{L}_{\mathrm{mix}}
 = -\frac{1}{\beta N}\sum_i \ln\Tr\,\mathcal{T}_\tau
   \exp\!\Big[-\!\int_0^{\beta}\! d\tau\,\hat H^{\mathrm{emb}}_{0,i}(\tau)\Big]
   \,,
\label{eq:Lmix}
\end{equation}
where
\begin{align}
\hat H_{0,i}^{\mathrm{emb}}(\tau)
 &= \sum_{ab}\big[\lambda_i(\tau)\big]_{ab} f^\dg_{ia} f_{ib}
  + \sum_{ab}\big[\lambda^c_i(\tau)\big]_{ab}\, b_{ib} b^\dg_{ia} \nonumber\\
 &\quad + \sum_{ab}\big(\big[\mathcal D_i(\tau) R_i^{T}(\tau)\big]_{ab} f^\dg_{ib} b_{ia}
   + \mathrm{H.c.}\big)
\label{eq:H0emb}
\end{align}
is the so-called non-interacting embedding Hamiltonian, recently introduced in Ref.~\cite{giuli2026}.

\begin{figure}[t]
\includegraphics[scale=0.41]{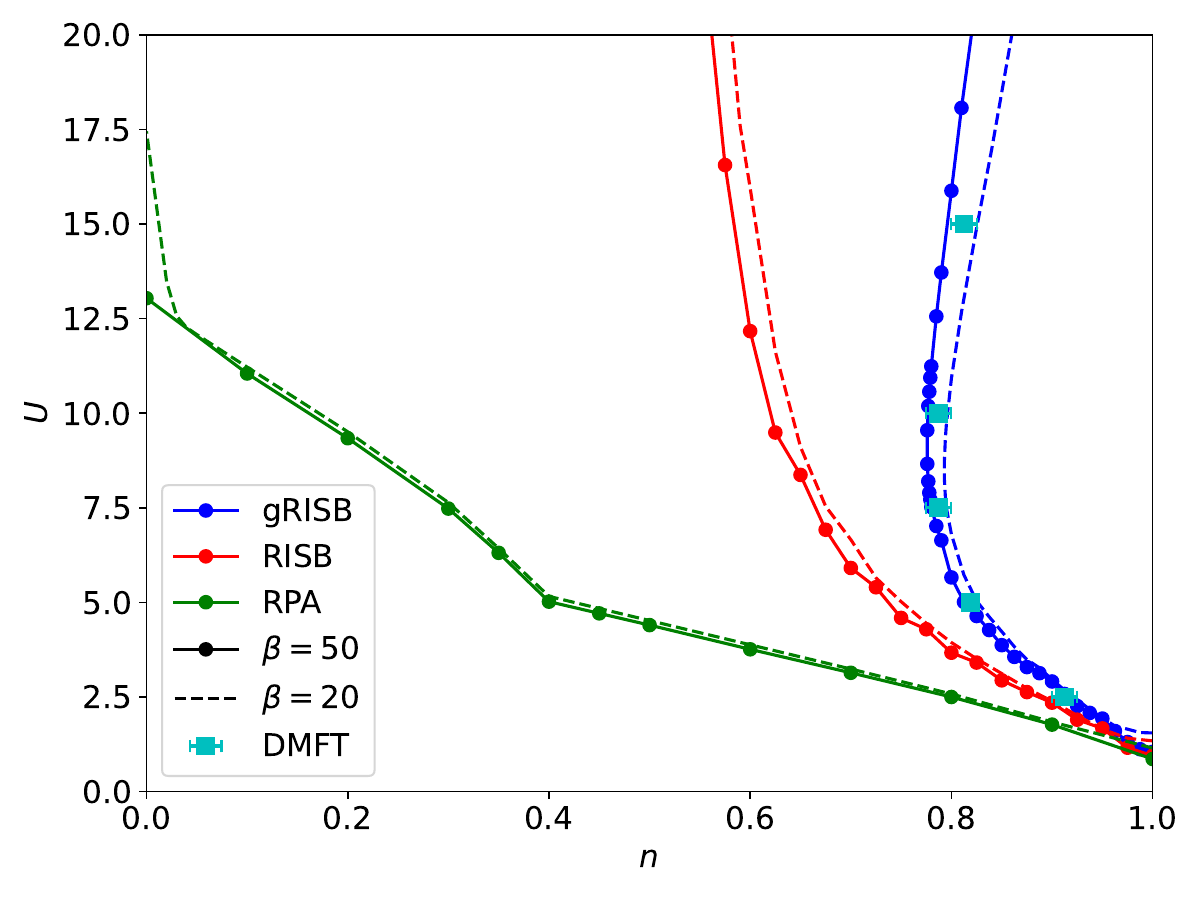}
\caption{ Magnetic phase boundary of the two-dimensional Hubbard model with nearest-neighbor hopping as a function
of electron filling $n$ and Coulomb interaction $U$ from the gRISB, standard RISB,
and RPA approaches at $\beta=50$ and $20$, and from DMFT at $\beta=50$.\label{fig:PD}}
\end{figure}

Our proposed linear-response formalism is obtained by expanding Eq.~\eqref{eq:dyn-lagrange} about the homogeneous, time-independent saddle point determined by $\partial_{\bm x}\mathcal{L}=0$, corresponding to the gRISB saddle-point solution.
The susceptibility then follows by adding a dynamical, spatially varying source coupled to the channel operator $\hat{\mathcal O}_i=\sum_{\sigma\sigma'}\mathcal O_{\sigma\sigma'}\,c^\dg_{i\sigma}c_{i\sigma'}$ [$\mathcal O=\sigma^0/\sqrt2$ (charge) or $\sigma^z/\sqrt2$ (spin)], $\Hemb_i(\tau) \to \Hemb_i(\tau) - \eta_{\bm q}(\tau)\, e^{i\bm q\cdot\bm R_i}\,\hat{\mathcal O}_i$, and differentiating $\mathcal{L}$ twice through the source-shifted saddle~\cite{Lee_PRX_2021,supplement}:
\begin{align}
&\chi_{\mathcal O\mathcal O}(\bm q,i\Om_n)
 = \chi^{\mathrm{emb}}_{\mathcal O\mathcal O}(i\Om_n)\nonumber \\
 &+ \sum_{\mu\nu}\chi^{\mathrm{emb}}_{\mathcal O x_\mu}(i\Om_n)\,
   \big[\mathcal{M}(\bm q,i\Om_n)\big]^{-1}_{\mu\nu}\,
   \chi^{\mathrm{emb}}_{x_\nu\mathcal O}(i\Om_n),
\label{eq:chi}
\end{align}
where $\chi^{\mathrm{emb}}_{\mathcal O\mathcal O}$ is the local Lehmann response of $\Hemb$ at the homogeneous $\eta=0$ saddle and $\chi^{\mathrm{emb}}_{\mathcal O x_\mu},\chi^{\mathrm{emb}}_{x_\nu\mathcal O}$ are the mixed responses coupling the operator to the field fluctuations $x_\mu$. Physically, the second term dresses the local response by the collective slave-boson fluctuations, whose propagator is the inverse fluctuation matrix
\begin{equation}
\big[\mathcal{M}(\mathbf{q},i\Omega_{n})\big]_{\mu\nu}=\frac{\partial^{2}\mathcal{L}}{\partial x_{\mu,-\mathbf{q},-\Omega_{n}}\partial x_{\nu,\mathbf{q},\Omega_{n}}}\big|_{\eta=0}.\label{eq:M}
\end{equation}
Here and below, the gauge zero modes of the static fluctuation matrix are projected out before the inversion~\cite{supplement}.
As the second variation of the dynamical Lagrangian, $\mathcal{M}(\bm q,i\Om_n)$ carries the full $(\bm q,i\Om_n)$ dependence: a dynamical Lindhard response of $\Hqp$ and $H^{\mathrm{emb}}_0$ plus the Lehmann representation of the interacting $\Hemb$.

\emph{Results}.--- Figure~\ref{fig:PD} shows the magnetic phase boundary determined from the divergence of the spin susceptibility within RISB, gRISB, RPA, and DMFT. RISB significantly corrects the overestimation of spin fluctuations in RPA, reducing the magnetic region from the entire filling range to fillings above $n\simeq0.6$, consistent with previous slave-boson studies~\cite{Riegler2020}. By extending the formalism to gRISB, the magnetic region is further reduced to fillings above $n\simeq0.8$. The gRISB phase boundary also closely tracks that of DMFT, which is calculated with the continuous-time quantum Monte Carlo impurity solver~\cite{Gull2011}.

\begin{figure}[t]
\begin{centering}
\includegraphics[scale=0.41]{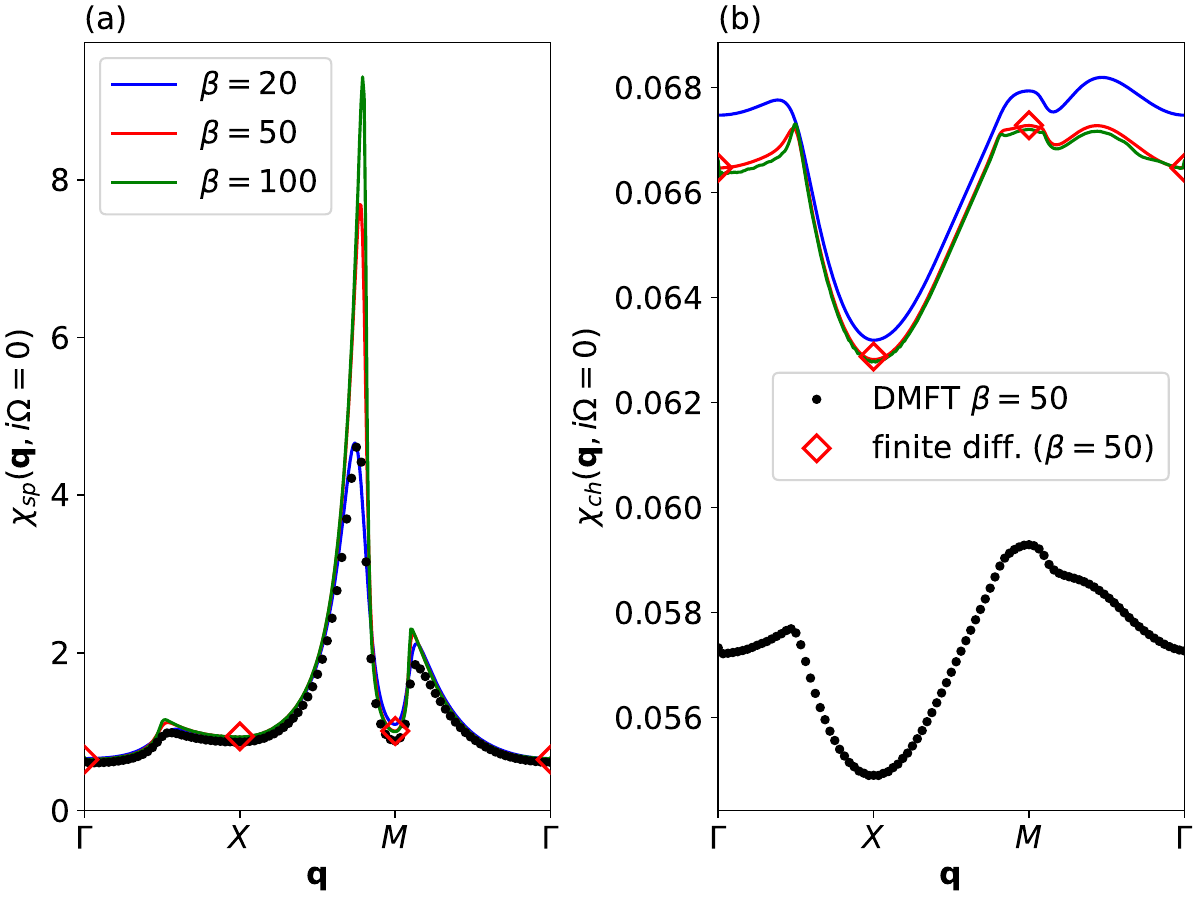}
\par\end{centering}
\caption{(a) The static spin susceptibility Re$\chi_{sp}(\mathbf{q},i\Omega_n{=}0)$ for gRISB with $\beta=20,\ 50, \ 100$ compared with DMFT at $\beta=50$. (b) The static charge susceptibility Re$\chi_{ch}(\mathbf{q},i\Omega_n{=}0)$ for gRISB with $\beta=20,\ 50, \ 100$ compared with DMFT at $\beta=50$. Here $U=10$ and $n=0.75$. The static susceptibilities computed from the finite-difference calculations at the commensurate momenta (red diamonds) are shown at $\beta=50$ for comparison. \label{fig:chi_static}}
\end{figure}

The gRISB static spin and charge susceptibilities along the high-symmetry path of the square-lattice Brillouin zone are shown in Fig.~\ref{fig:chi_static} for several temperatures at electron filling $n=0.75$ and $U=10$ near the magnetic phase boundary. The spin susceptibility, Fig.~\ref{fig:chi_static}(a), shows a sharp peak at the incommensurate momentum around $\mathbf{q}=(\pi,0.7\pi)$, indicating the proximity to an incommensurate magnetic instability. Lowering the temperature further enhances the spin susceptibility towards the instability. The gRISB spin susceptibility captures the momentum dependence, although the peak magnitude differs from DMFT. The static charge susceptibility [Fig.~\ref{fig:chi_static}(b)] is suppressed to low values due to the strong Coulomb interaction, where the $X$ point shows the strongest suppression. The DMFT charge susceptibility shows a similar behavior. The remaining discrepancy primarily reflects the finite auxiliary-bath truncation and the approximate finite-bath gRISB self-energy. We also validate our static susceptibilities by comparing them with the supercell saddle-point finite-difference calculations at the commensurate momenta (red diamonds), which yield consistent values.

\begin{figure}[t]
\begin{centering}
\includegraphics[scale=0.41]{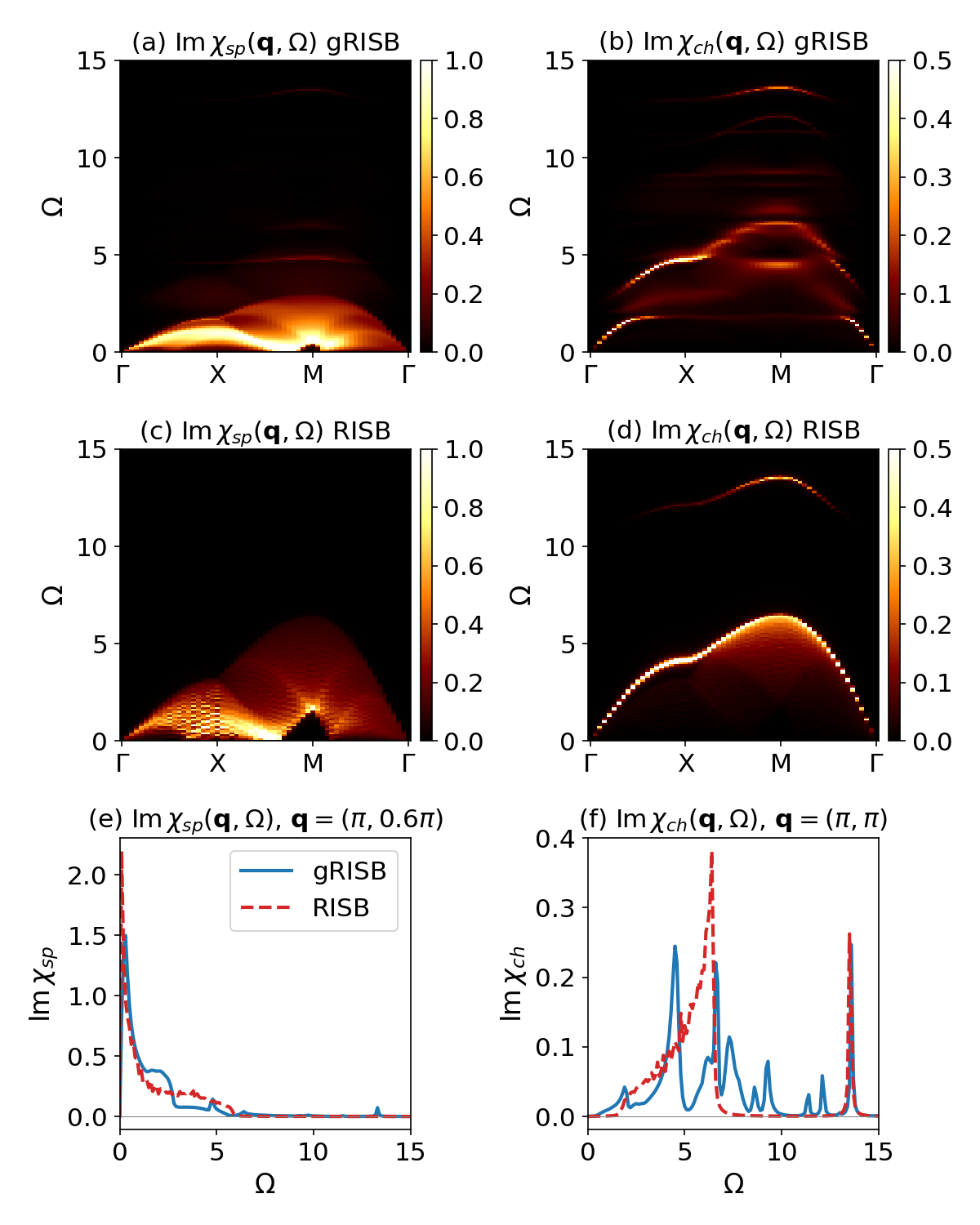}
\par\end{centering}
\caption{(a) The dynamical spin susceptibility on the real-frequency axis Im$\chi_{sp}(\mathbf{q},\Omega)$ for gRISB with $\beta=50$. (b) The dynamical charge susceptibility Im$\chi_{ch}(\mathbf{q},\Omega)$
for gRISB with $\beta=50$. (c) The dynamical spin susceptibility Im$\chi_{sp}(\mathbf{q},\Omega)$ for RISB with $\beta=50$. (d) The dynamical charge susceptibility Im$\chi_{ch}(\mathbf{q},\Omega)$ for RISB with $\beta=50$.  (e) The gRISB and RISB dynamical spin susceptibility at $\mathbf{q}=(\pi,0.6\pi)$. (f) The gRISB and RISB charge susceptibility at $\mathbf{q}=(\pi,\pi)$. Here $U=10$, and the filling is $n=0.75$ and $n=0.6$ for gRISB and RISB, respectively.\label{fig:chi_dyn}}
\end{figure}

Figure~\ref{fig:chi_dyn}(a) shows the real-frequency gRISB dynamical spin susceptibility $\mathrm{Im}\chi_{sp}(\mathbf{q},\Omega)$ at $n=0.75$, $U=10$, and $\beta=50$. It exhibits a low-energy dispersive mode between $\Omega=0$ and $\Omega=2.5$, the paramagnon excitation, peaked strongly around the incommensurate momentum $\mathbf{q}=(\pi,0.7\pi)$, on top of the particle--hole continuum. The dynamical gRISB charge susceptibility $\mathrm{Im}\chi_{ch}(\mathbf{q},\Omega)$, shown in Fig.~\ref{fig:chi_dyn}(b), reveals rich excitations. The low-energy dispersive sound modes stemming from the $\Gamma$ point show richer behavior compared to the standard RISB charge susceptibility, Fig.~\ref{fig:chi_dyn}(d). The high-energy Hubbard excitation around $\Omega=12$ is also well captured, and several satellite charge excitations appear between the sound modes and the Hubbard excitations. The RISB spin and charge susceptibilities [Fig.~\ref{fig:chi_dyn}(c),(d), $n=0.6$] show a wider paramagnon bandwidth [Fig.~\ref{fig:chi_dyn}(e)] and fewer charge excitations [Fig.~\ref{fig:chi_dyn}(f)] than gRISB, reflecting the weaker renormalization ($Z_{\mathrm{RISB}}=0.81$ vs.\ $Z_{\mathrm{gRISB}}=0.45$); the KRSB sound modes and Hubbard excitation are well reproduced~\cite{Dao2017,Riegler2020,supplement}.

\emph{Conserving quasiparticle approximation.}--- To obtain a
diagrammatic representation of the response in terms of Green's
functions and vertex corrections, we introduce a conserving
quasiparticle approximation (CQPA), extending the quasiparticle
approximation of Ref.~\cite{Lee_PRX_2021} to gRISB.
Following Ref.~\cite{giuli2026}, we consider the operators
\begin{align}
\tilde c_{i\sigma}
&=
\sum_a[R_i^\dg]_{\sigma a}\,f_{ia}
\,.
\label{eq:ctilde}
\end{align}
Then, we couple the external source to
$\sum_{\sigma\sigma'}\mathcal O_{\sigma\sigma'}\,
\tilde c^\dg_{i\sigma}\tilde c_{i\sigma'}$
in the quasiparticle Hamiltonian.
This choice is motivated by the equilibrium complete-bath limit,
where, as proven in Ref.~\cite{giuli2026}, $R_i^\dg R_i\to 1$
and the quasiparticle and embedding evaluations of the local
one-body density matrix agree.
At finite bath size, we impose the corresponding matching of the
channel densities as a response constraint, through a dynamical Lagrange-multiplier field
$\zeta_{\bm q,\Omega_n}$ for each channel:
\begin{align}
\Big\langle
\sum_{\sigma\sigma'}\mathcal O_{\sigma\sigma'}\,
\tilde c^\dg_{i\sigma}\tilde c_{i\sigma'}
\Big\rangle_{\hat{H}^{\rm qp}}
&=
\big\langle\hat{\mathcal O}_i\big\rangle_{\Hemb}
\,.
\label{eq:cqpaconstraint}
\end{align}
At the unperturbed saddle this equality holds only approximately, because the
quasiparticle representation misses part of the spectral weight; the mismatch vanishes
in the complete-bath limit~\cite{supplement}. We therefore keep $\zeta=0$ at
the saddle and impose Eq.~\eqref{eq:cqpaconstraint} at the Gaussian level, including
$\delta\zeta$ together with the fluctuations of $\bm x$.

\begin{figure}[t]
\begin{centering}
\includegraphics[scale=0.42]{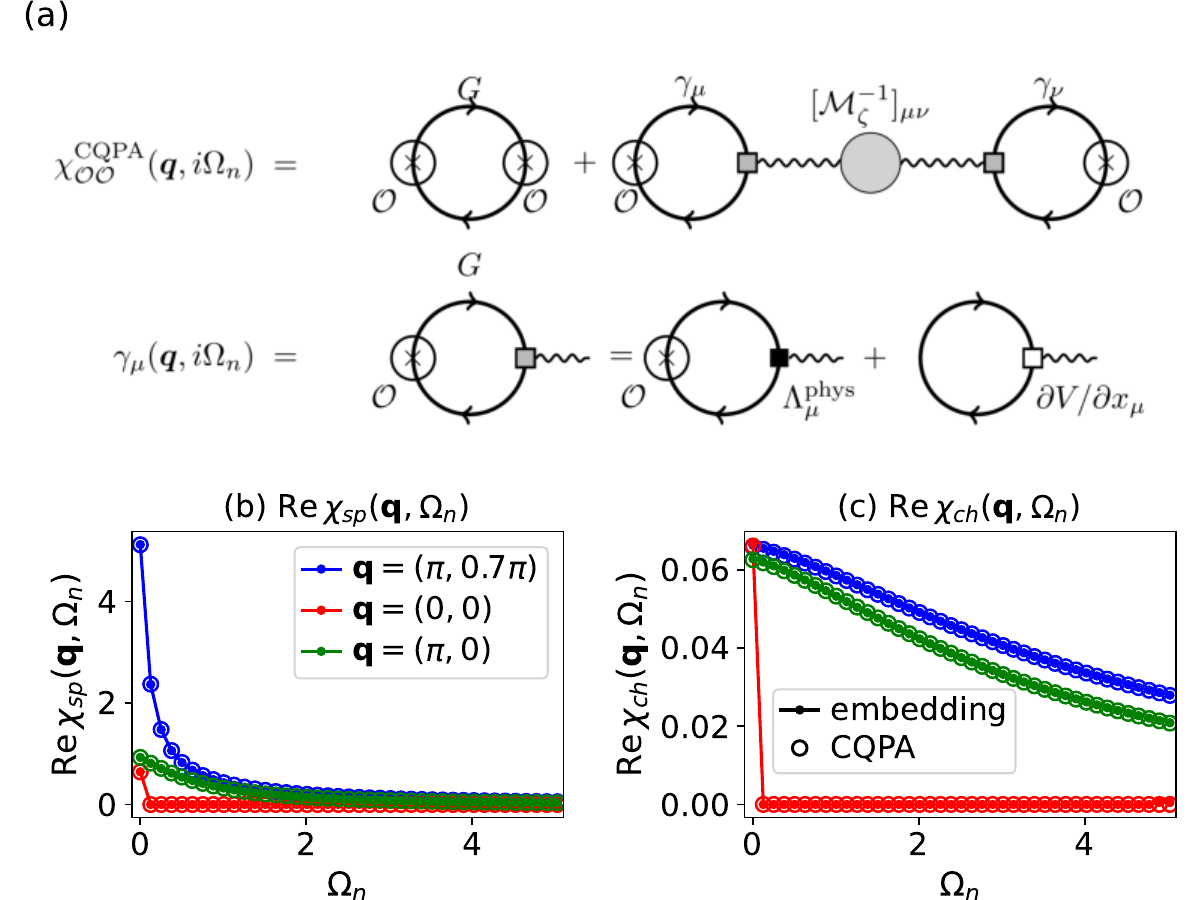}
\par\end{centering}
\caption{(a) Diagrammatic representation of the susceptibility within the conserving quasiparticle approximation (CQPA). See main text for details. (b) The CQPA dynamical spin susceptibility on the Matsubara axis Re$\chi_{sp}(\mathbf{q},i\Omega_n)$ for gRISB and selected momenta $\mathbf{q}$ compared with the embedding formalism [Eq.~\eqref{eq:chi}]. (c) The CQPA dynamical charge susceptibility on the Matsubara axis Re$\chi_{ch}(\mathbf{q},i\Omega_n)$ for gRISB and selected momenta $\mathbf{q}$ compared with the embedding formalism. Here $U=10$, $n=0.75$, and $\beta=50$.\label{fig:CQPA}}
\end{figure}

The CQPA susceptibility admits the ``bubble + vertex--boson--vertex''
representation of Fig.~\ref{fig:CQPA}(a). Using the four-vector notation
$k\equiv(\bm k,i\omega_n)$, $q\equiv(\bm q,i\Omega_n)$, and writing
$G_k = R^\dagger\underline G_k R$ for the physical Green's function
[$\underline G_k=(i\omega_n-\Hqp_{\bm k})^{-1}$ the
quasiparticle one], we have
\begin{align}
&\chi_{\mathcal O\mathcal O}(q)
 = \chi^{(0)}(q)
 + \textstyle\sum_{\mu\nu}\gamma_\mu(-q)\,
   \big[\tilde{\mathcal M}_\zeta(q)\big]^{-1}_{\mu\nu}\,\gamma_\nu(q) ,
\label{eq:cqpa}\\
&\chi^{(0)}(q)
 = -\tfrac{1}{\beta N}\textstyle\sum_{k}
 \mathrm{Tr}\big[\mathcal O\,G_{k+q}\,\mathcal O\,G_k\big],
\label{eq:chi0}\\
&\gamma_\mu(q)
 = -\tfrac{1}{\beta N}\textstyle\sum_{k}
 \mathrm{Tr}\big(\mathcal O\,G_k\,\Lambda^{\mathrm{phys}}_\mu(k,q)\,G_{k+q}\big)
\nonumber\\
&\hspace{2.6em}
 -\tfrac{1}{N}\textstyle\sum_{\bm k}
 \mathrm{tr}\big(\nF(\Hqp_{\bm k})\,\partial V/\partial x_\mu\big).
\label{eq:triangle}
\end{align}
Here $\Lambda^{\mathrm{phys}}_\mu(k,q)
 = G_k^{-1}\big(R^\dagger\underline G_k\,\Lambda_\mu(\bm k,\bm q)\,
 \underline G_{k+q}R\big)G_{k+q}^{-1}$ is the amputated physical vertex [solid square in Fig.~\ref{fig:CQPA}(a)]
built from the bare quasiparticle vertex
$\Lambda_\mu(\bm k,\bm q)=\partial[\underline H^{\mathrm{qp}}]_{\bm k,\bm k+\bm q}
 /\partial x_{\mu,-\bm q}$. The second term [open square in
Fig.~\ref{fig:CQPA}(a)] is the partial derivative of the dressed vertex
$V=R\,\mathcal O R^\dagger$ weighted by the (frequency-independent) quasiparticle
occupation $\nF(\Hqp_{\bm k})$ ($\nF$ the Fermi function), a diamagnetic-like contact
required for gauge covariance of the physical vertex. The boson propagator
$\big[\tilde{\mathcal M}_\zeta(q)\big]^{-1}$ is the pseudo-inverse on the physical
subspace~\cite{supplement} of the fluctuation matrix of Eq.~\eqref{eq:M} in the
$\zeta$-enlarged space $\bm x\to(\bm x,\zeta)$, with $\mu,\nu$ in
Eq.~\eqref{eq:cqpa} running over $(\bm x,\zeta)$,
$\gamma_\zeta(q)=\chi^{(0)}(q)$, and $\gamma_\mu=0$ on the embedding
directions $(D',D'',\lambda^c)$.
The $\zeta$ constraint enforces the conservation law $\chi(\bm q{=}0,i\Omega_n{\neq}0)=0$.

The CQPA results are shown in Fig.~\ref{fig:CQPA}(b) and (c) for the gRISB spin and charge susceptibilities, respectively, on the Matsubara axis for selected momenta. The CQPA agrees with Eq.~\eqref{eq:chi} to within $1\%$ of the peak value at all sampled momenta and Matsubara frequencies~\cite{supplement}. On the real-frequency axis, the CQPA and the embedding susceptibility [Eq.~\eqref{eq:chi}] differ in their line shapes, while the major excitations remain qualitatively captured~\cite{supplement}.

\emph{Conclusion}.--- We proposed a finite-temperature dynamical extension of gRISB and used its
Gaussian fluctuations to compute the response functions of the two-dimensional Hubbard model. The responses closely track the DMFT
magnetic phase boundary and capture the qualitative trends of the static susceptibilities while markedly improving upon RISB and RPA,
and recover the Kotliar--Ruckenstein response at the minimal RISB level, thereby bridging the KRSB
and DMFT limits. Evaluated analytically on the real-frequency axis, the dynamical susceptibilities
capture the paramagnon, the sound modes, and the high-energy Hubbard excitations, free from
sampling and analytic-continuation error and directly comparable to the dynamical structure factors measured
by inelastic neutron scattering and resonant inelastic X-ray scattering.
The approach also admits a diagrammatic representation in dressed
Green's functions and vertex corrections, which provides a natural starting point for
systematic extensions. As gRISB is already an efficient multiorbital ab initio
method, the present formalism opens a practical route to the real-frequency two-particle response of
correlated materials. A rigorous derivation of Eq.~\eqref{eq:dyn-lagrange}, along the routes outlined above,
remains an important direction.

\emph{Acknowledgments}.---
T.-H.L. gratefully acknowledges funding from the National Science and Technology Council (NSTC) of Taiwan under Grant Nos. NSTC 112-2112-M-194-007-MY3 and 115-2112-M-194-005, as well as support from the National Center for Theoretical Sciences (NCTS) in Taiwan.
C.M. and G.K. were supported by the US Department
of Energy, Office of Basic Energy Sciences, as part of the
Computational Materials Science Program.
N.L. gratefully acknowledges funding from the National Science Foundation under Award No. DMR-2532771 and from the Simons Foundation (Grant No. 00024037).

\bibliographystyle{apsrev}
\bibliography{ref}

\begin{thebibliography}{38}
\expandafter\ifx\csname natexlab\endcsname\relax\def\natexlab#1{#1}\fi
\expandafter\ifx\csname bibnamefont\endcsname\relax
  \def\bibnamefont#1{#1}\fi
\expandafter\ifx\csname bibfnamefont\endcsname\relax
  \def\bibfnamefont#1{#1}\fi
\expandafter\ifx\csname citenamefont\endcsname\relax
  \def\citenamefont#1{#1}\fi
\expandafter\ifx\csname url\endcsname\relax
  \def\url#1{\texttt{#1}}\fi
\expandafter\ifx\csname urlprefix\endcsname\relax\def\urlprefix{URL }\fi
\providecommand{\bibinfo}[2]{#2}
\providecommand{\eprint}[2][]{\url{#2}}

\bibitem[{\citenamefont{Moriya}(1985)}]{Moriya1985}
\bibinfo{author}{\bibfnamefont{T.}~\bibnamefont{Moriya}},
  \emph{\bibinfo{title}{Spin Fluctuations in Itinerant Electron Magnetism}},
  vol.~\bibinfo{volume}{56} of \emph{\bibinfo{series}{Springer Series in
  Solid-State Sciences}} (\bibinfo{publisher}{Springer-Verlag},
  \bibinfo{address}{Berlin, Heidelberg}, \bibinfo{year}{1985}), ISBN
  \bibinfo{isbn}{978-0-387-15422-0}.

\bibitem[{\citenamefont{Sch\"afer et~al.}(2021)\citenamefont{Sch\"afer,
  Wentzell, \ifmmode~\check{S}\else \v{S}\fi{}imkovic, He, Hille, Klett,
  Eckhardt, Arzhang, Harkov, Le~R{\'e}gent et~al.}}]{Schafer2021}
\bibinfo{author}{\bibfnamefont{T.}~\bibnamefont{Sch\"afer}},
  \bibinfo{author}{\bibfnamefont{N.}~\bibnamefont{Wentzell}},
  \bibinfo{author}{\bibfnamefont{F.}~\bibnamefont{\ifmmode~\check{S}\else
  \v{S}\fi{}imkovic}}, \bibinfo{author}{\bibfnamefont{Y.-Y.} \bibnamefont{He}},
  \bibinfo{author}{\bibfnamefont{C.}~\bibnamefont{Hille}},
  \bibinfo{author}{\bibfnamefont{M.}~\bibnamefont{Klett}},
  \bibinfo{author}{\bibfnamefont{C.~J.} \bibnamefont{Eckhardt}},
  \bibinfo{author}{\bibfnamefont{B.}~\bibnamefont{Arzhang}},
  \bibinfo{author}{\bibfnamefont{V.}~\bibnamefont{Harkov}},
  \bibinfo{author}{\bibfnamefont{F.-M.} \bibnamefont{Le~R{\'e}gent}},
  \bibnamefont{et~al.}, \bibinfo{journal}{Phys. Rev. X}
  \textbf{\bibinfo{volume}{11}}, \bibinfo{pages}{011058}
  (\bibinfo{year}{2021}),
  \urlprefix\url{https://link.aps.org/doi/10.1103/PhysRevX.11.011058}.

\bibitem[{\citenamefont{Bickers et~al.}(1989)\citenamefont{Bickers, Scalapino,
  and White}}]{Bickers1989}
\bibinfo{author}{\bibfnamefont{N.~E.} \bibnamefont{Bickers}},
  \bibinfo{author}{\bibfnamefont{D.~J.} \bibnamefont{Scalapino}},
  \bibnamefont{and} \bibinfo{author}{\bibfnamefont{S.~R.} \bibnamefont{White}},
  \bibinfo{journal}{Phys. Rev. Lett.} \textbf{\bibinfo{volume}{62}},
  \bibinfo{pages}{961} (\bibinfo{year}{1989}),
  \urlprefix\url{https://link.aps.org/doi/10.1103/PhysRevLett.62.961}.

\bibitem[{\citenamefont{Vilk and Tremblay}(1997)}]{Vilk1997}
\bibinfo{author}{\bibfnamefont{Y.~M.} \bibnamefont{Vilk}} \bibnamefont{and}
  \bibinfo{author}{\bibfnamefont{A.-M.~S.} \bibnamefont{Tremblay}},
  \bibinfo{journal}{Journal de Physique I France} \textbf{\bibinfo{volume}{7}},
  \bibinfo{pages}{1309} (\bibinfo{year}{1997}).

\bibitem[{\citenamefont{Georges et~al.}(1996)\citenamefont{Georges, Kotliar,
  Krauth, and Rozenberg}}]{DMFT_RMP_1996}
\bibinfo{author}{\bibfnamefont{A.}~\bibnamefont{Georges}},
  \bibinfo{author}{\bibfnamefont{G.}~\bibnamefont{Kotliar}},
  \bibinfo{author}{\bibfnamefont{W.}~\bibnamefont{Krauth}}, \bibnamefont{and}
  \bibinfo{author}{\bibfnamefont{M.~J.} \bibnamefont{Rozenberg}},
  \bibinfo{journal}{Rev. Mod. Phys.} \textbf{\bibinfo{volume}{68}},
  \bibinfo{pages}{13} (\bibinfo{year}{1996}),
  \urlprefix\url{https://link.aps.org/doi/10.1103/RevModPhys.68.13}.

\bibitem[{\citenamefont{Kotliar et~al.}(2006)\citenamefont{Kotliar, Savrasov,
  Haule, Oudovenko, Parcollet, and Marianetti}}]{DMFT_RMP_2006}
\bibinfo{author}{\bibfnamefont{G.}~\bibnamefont{Kotliar}},
  \bibinfo{author}{\bibfnamefont{S.~Y.} \bibnamefont{Savrasov}},
  \bibinfo{author}{\bibfnamefont{K.}~\bibnamefont{Haule}},
  \bibinfo{author}{\bibfnamefont{V.~S.} \bibnamefont{Oudovenko}},
  \bibinfo{author}{\bibfnamefont{O.}~\bibnamefont{Parcollet}},
  \bibnamefont{and} \bibinfo{author}{\bibfnamefont{C.~A.}
  \bibnamefont{Marianetti}}, \bibinfo{journal}{Rev. Mod. Phys.}
  \textbf{\bibinfo{volume}{78}}, \bibinfo{pages}{865} (\bibinfo{year}{2006}),
  \urlprefix\url{https://link.aps.org/doi/10.1103/RevModPhys.78.865}.

\bibitem[{\citenamefont{Rohringer et~al.}(2018)\citenamefont{Rohringer,
  Hafermann, Toschi, Katanin, Antipov, Katsnelson, Lichtenstein, Rubtsov, and
  Held}}]{Rohringer2018}
\bibinfo{author}{\bibfnamefont{G.}~\bibnamefont{Rohringer}},
  \bibinfo{author}{\bibfnamefont{H.}~\bibnamefont{Hafermann}},
  \bibinfo{author}{\bibfnamefont{A.}~\bibnamefont{Toschi}},
  \bibinfo{author}{\bibfnamefont{A.~A.} \bibnamefont{Katanin}},
  \bibinfo{author}{\bibfnamefont{A.~E.} \bibnamefont{Antipov}},
  \bibinfo{author}{\bibfnamefont{M.~I.} \bibnamefont{Katsnelson}},
  \bibinfo{author}{\bibfnamefont{A.~I.} \bibnamefont{Lichtenstein}},
  \bibinfo{author}{\bibfnamefont{A.~N.} \bibnamefont{Rubtsov}},
  \bibnamefont{and} \bibinfo{author}{\bibfnamefont{K.}~\bibnamefont{Held}},
  \bibinfo{journal}{Rev. Mod. Phys.} \textbf{\bibinfo{volume}{90}},
  \bibinfo{pages}{025003} (\bibinfo{year}{2018}),
  \urlprefix\url{https://link.aps.org/doi/10.1103/RevModPhys.90.025003}.

\bibitem[{\citenamefont{Gull et~al.}(2011)\citenamefont{Gull, Millis,
  Lichtenstein, Rubtsov, Troyer, and Werner}}]{Gull2011}
\bibinfo{author}{\bibfnamefont{E.}~\bibnamefont{Gull}},
  \bibinfo{author}{\bibfnamefont{A.~J.} \bibnamefont{Millis}},
  \bibinfo{author}{\bibfnamefont{A.~I.} \bibnamefont{Lichtenstein}},
  \bibinfo{author}{\bibfnamefont{A.~N.} \bibnamefont{Rubtsov}},
  \bibinfo{author}{\bibfnamefont{M.}~\bibnamefont{Troyer}}, \bibnamefont{and}
  \bibinfo{author}{\bibfnamefont{P.}~\bibnamefont{Werner}},
  \bibinfo{journal}{Rev. Mod. Phys.} \textbf{\bibinfo{volume}{83}},
  \bibinfo{pages}{349} (\bibinfo{year}{2011}),
  \urlprefix\url{https://link.aps.org/doi/10.1103/RevModPhys.83.349}.

\bibitem[{\citenamefont{Lechermann et~al.}(2007)\citenamefont{Lechermann,
  Georges, Kotliar, and Parcollet}}]{Lecherman_2007}
\bibinfo{author}{\bibfnamefont{F.}~\bibnamefont{Lechermann}},
  \bibinfo{author}{\bibfnamefont{A.}~\bibnamefont{Georges}},
  \bibinfo{author}{\bibfnamefont{G.}~\bibnamefont{Kotliar}}, \bibnamefont{and}
  \bibinfo{author}{\bibfnamefont{O.}~\bibnamefont{Parcollet}},
  \bibinfo{journal}{Phys. Rev. B} \textbf{\bibinfo{volume}{76}},
  \bibinfo{pages}{155102} (\bibinfo{year}{2007}),
  \urlprefix\url{https://link.aps.org/doi/10.1103/PhysRevB.76.155102}.

\bibitem[{\citenamefont{B\"unemann and Gebhard}(2007)}]{Bunemann_2007}
\bibinfo{author}{\bibfnamefont{J.}~\bibnamefont{B\"unemann}} \bibnamefont{and}
  \bibinfo{author}{\bibfnamefont{F.}~\bibnamefont{Gebhard}},
  \bibinfo{journal}{Phys. Rev. B} \textbf{\bibinfo{volume}{76}},
  \bibinfo{pages}{193104} (\bibinfo{year}{2007}),
  \urlprefix\url{https://link.aps.org/doi/10.1103/PhysRevB.76.193104}.

\bibitem[{\citenamefont{Lanat\`a et~al.}(2015)\citenamefont{Lanat\`a, Yao,
  Wang, Ho, and Kotliar}}]{Lanata_2015_PRX}
\bibinfo{author}{\bibfnamefont{N.}~\bibnamefont{Lanat\`a}},
  \bibinfo{author}{\bibfnamefont{Y.}~\bibnamefont{Yao}},
  \bibinfo{author}{\bibfnamefont{C.-Z.} \bibnamefont{Wang}},
  \bibinfo{author}{\bibfnamefont{K.-M.} \bibnamefont{Ho}}, \bibnamefont{and}
  \bibinfo{author}{\bibfnamefont{G.}~\bibnamefont{Kotliar}},
  \bibinfo{journal}{Phys. Rev. X} \textbf{\bibinfo{volume}{5}},
  \bibinfo{pages}{011008} (\bibinfo{year}{2015}),
  \urlprefix\url{https://link.aps.org/doi/10.1103/PhysRevX.5.011008}.

\bibitem[{\citenamefont{Lanat\`a
  et~al.}(2017{\natexlab{a}})\citenamefont{Lanat\`a, Yao, Deng,
  Dobrosavljevi\ifmmode~\acute{c}\else \'{c}\fi{}, and
  Kotliar}}]{Lanata_2017_PRL}
\bibinfo{author}{\bibfnamefont{N.}~\bibnamefont{Lanat\`a}},
  \bibinfo{author}{\bibfnamefont{Y.}~\bibnamefont{Yao}},
  \bibinfo{author}{\bibfnamefont{X.}~\bibnamefont{Deng}},
  \bibinfo{author}{\bibfnamefont{V.}~\bibnamefont{Dobrosavljevi\ifmmode~\acute{c}\else
  \'{c}\fi{}}}, \bibnamefont{and}
  \bibinfo{author}{\bibfnamefont{G.}~\bibnamefont{Kotliar}},
  \bibinfo{journal}{Phys. Rev. Lett.} \textbf{\bibinfo{volume}{118}},
  \bibinfo{pages}{126401} (\bibinfo{year}{2017}{\natexlab{a}}),
  \urlprefix\url{https://link.aps.org/doi/10.1103/PhysRevLett.118.126401}.

\bibitem[{\citenamefont{Seibold and Lorenzana}(2001)}]{Seibold2001_charge}
\bibinfo{author}{\bibfnamefont{G.}~\bibnamefont{Seibold}} \bibnamefont{and}
  \bibinfo{author}{\bibfnamefont{J.}~\bibnamefont{Lorenzana}},
  \bibinfo{journal}{Phys. Rev. Lett.} \textbf{\bibinfo{volume}{86}},
  \bibinfo{pages}{2605} (\bibinfo{year}{2001}),
  \urlprefix\url{https://link.aps.org/doi/10.1103/PhysRevLett.86.2605}.

\bibitem[{\citenamefont{Seibold et~al.}(2004)\citenamefont{Seibold, Becca,
  Rubin, and Lorenzana}}]{Seibold2004_magnetic}
\bibinfo{author}{\bibfnamefont{G.}~\bibnamefont{Seibold}},
  \bibinfo{author}{\bibfnamefont{F.}~\bibnamefont{Becca}},
  \bibinfo{author}{\bibfnamefont{P.}~\bibnamefont{Rubin}}, \bibnamefont{and}
  \bibinfo{author}{\bibfnamefont{J.}~\bibnamefont{Lorenzana}},
  \bibinfo{journal}{Phys. Rev. B} \textbf{\bibinfo{volume}{69}},
  \bibinfo{pages}{155113} (\bibinfo{year}{2004}),
  \urlprefix\url{https://link.aps.org/doi/10.1103/PhysRevB.69.155113}.

\bibitem[{\citenamefont{Oelsen et~al.}(2011)\citenamefont{Oelsen, Seibold, and
  B\"unemann}}]{von_Oelsen_2011_PRL}
\bibinfo{author}{\bibfnamefont{E.~v.} \bibnamefont{Oelsen}},
  \bibinfo{author}{\bibfnamefont{G.}~\bibnamefont{Seibold}}, \bibnamefont{and}
  \bibinfo{author}{\bibfnamefont{J.}~\bibnamefont{B\"unemann}},
  \bibinfo{journal}{Phys. Rev. Lett.} \textbf{\bibinfo{volume}{107}},
  \bibinfo{pages}{076402} (\bibinfo{year}{2011}),
  \urlprefix\url{https://link.aps.org/doi/10.1103/PhysRevLett.107.076402}.

\bibitem[{\citenamefont{Lee et~al.}(2021)\citenamefont{Lee, Lanat\`a, Kim, and
  Kotliar}}]{Lee_PRX_2021}
\bibinfo{author}{\bibfnamefont{T.-H.} \bibnamefont{Lee}},
  \bibinfo{author}{\bibfnamefont{N.}~\bibnamefont{Lanat\`a}},
  \bibinfo{author}{\bibfnamefont{M.}~\bibnamefont{Kim}}, \bibnamefont{and}
  \bibinfo{author}{\bibfnamefont{G.}~\bibnamefont{Kotliar}},
  \bibinfo{journal}{Phys. Rev. X} \textbf{\bibinfo{volume}{11}},
  \bibinfo{pages}{041040} (\bibinfo{year}{2021}),
  \urlprefix\url{https://link.aps.org/doi/10.1103/PhysRevX.11.041040}.

\bibitem[{\citenamefont{Kotliar and Ruckenstein}(1986)}]{Kotliar1986}
\bibinfo{author}{\bibfnamefont{G.}~\bibnamefont{Kotliar}} \bibnamefont{and}
  \bibinfo{author}{\bibfnamefont{A.~E.} \bibnamefont{Ruckenstein}},
  \bibinfo{journal}{Phys. Rev. Lett.} \textbf{\bibinfo{volume}{57}},
  \bibinfo{pages}{1362} (\bibinfo{year}{1986}),
  \urlprefix\url{https://link.aps.org/doi/10.1103/PhysRevLett.57.1362}.

\bibitem[{\citenamefont{Lavagna}(1990)}]{Lavagna1990}
\bibinfo{author}{\bibfnamefont{M.}~\bibnamefont{Lavagna}},
  \bibinfo{journal}{Phys. Rev. B} \textbf{\bibinfo{volume}{41}},
  \bibinfo{pages}{142} (\bibinfo{year}{1990}),
  \urlprefix\url{https://link.aps.org/doi/10.1103/PhysRevB.41.142}.

\bibitem[{\citenamefont{Li et~al.}(1989)\citenamefont{Li, W\"olfle, and
  Hirschfeld}}]{Li_1989_PRB_SRISB}
\bibinfo{author}{\bibfnamefont{T.}~\bibnamefont{Li}},
  \bibinfo{author}{\bibfnamefont{P.}~\bibnamefont{W\"olfle}}, \bibnamefont{and}
  \bibinfo{author}{\bibfnamefont{P.~J.} \bibnamefont{Hirschfeld}},
  \bibinfo{journal}{Phys. Rev. B} \textbf{\bibinfo{volume}{40}},
  \bibinfo{pages}{6817} (\bibinfo{year}{1989}),
  \urlprefix\url{https://link.aps.org/doi/10.1103/PhysRevB.40.6817}.

\bibitem[{\citenamefont{Li et~al.}(1991)\citenamefont{Li, Sun, and
  W{\"o}lfle}}]{Li1991}
\bibinfo{author}{\bibfnamefont{T.}~\bibnamefont{Li}},
  \bibinfo{author}{\bibfnamefont{Y.~S.} \bibnamefont{Sun}}, \bibnamefont{and}
  \bibinfo{author}{\bibfnamefont{P.}~\bibnamefont{W{\"o}lfle}},
  \bibinfo{journal}{Zeitschrift f{\"u}r Physik B Condensed Matter}
  \textbf{\bibinfo{volume}{82}}, \bibinfo{pages}{369} (\bibinfo{year}{1991}),
  ISSN \bibinfo{issn}{1431-584X},
  \urlprefix\url{https://doi.org/10.1007/BF01357181}.

\bibitem[{\citenamefont{Zimmermann et~al.}(1997)\citenamefont{Zimmermann,
  Fr\'esard, and W\"olfle}}]{Zimmermann1997}
\bibinfo{author}{\bibfnamefont{W.}~\bibnamefont{Zimmermann}},
  \bibinfo{author}{\bibfnamefont{R.}~\bibnamefont{Fr\'esard}},
  \bibnamefont{and} \bibinfo{author}{\bibfnamefont{P.}~\bibnamefont{W\"olfle}},
  \bibinfo{journal}{Phys. Rev. B} \textbf{\bibinfo{volume}{56}},
  \bibinfo{pages}{10097} (\bibinfo{year}{1997}),
  \urlprefix\url{https://link.aps.org/doi/10.1103/PhysRevB.56.10097}.

\bibitem[{\citenamefont{Dao and Fr\'esard}(2017)}]{Dao2017}
\bibinfo{author}{\bibfnamefont{V.~H.} \bibnamefont{Dao}} \bibnamefont{and}
  \bibinfo{author}{\bibfnamefont{R.}~\bibnamefont{Fr\'esard}},
  \bibinfo{journal}{Phys. Rev. B} \textbf{\bibinfo{volume}{95}},
  \bibinfo{pages}{165127} (\bibinfo{year}{2017}),
  \urlprefix\url{https://link.aps.org/doi/10.1103/PhysRevB.95.165127}.

\bibitem[{\citenamefont{Lanat\`a
  et~al.}(2017{\natexlab{b}})\citenamefont{Lanat\`a, Lee, Yao, and
  Dobrosavljevi\ifmmode~\acute{c}\else \'{c}\fi{}}}]{gRISB_2017}
\bibinfo{author}{\bibfnamefont{N.}~\bibnamefont{Lanat\`a}},
  \bibinfo{author}{\bibfnamefont{T.-H.} \bibnamefont{Lee}},
  \bibinfo{author}{\bibfnamefont{Y.-X.} \bibnamefont{Yao}}, \bibnamefont{and}
  \bibinfo{author}{\bibfnamefont{V.}~\bibnamefont{Dobrosavljevi\ifmmode~\acute{c}\else
  \'{c}\fi{}}}, \bibinfo{journal}{Phys. Rev. B} \textbf{\bibinfo{volume}{96}},
  \bibinfo{pages}{195126} (\bibinfo{year}{2017}{\natexlab{b}}),
  \urlprefix\url{https://link.aps.org/doi/10.1103/PhysRevB.96.195126}.

\bibitem[{\citenamefont{Frank et~al.}(2021)\citenamefont{Frank, Lee,
  Bhattacharyya, Tsang, Quito, Dobrosavljevi\ifmmode~\acute{c}\else \'{c}\fi{},
  Christiansen, and Lanat\`a}}]{gRISB_2021}
\bibinfo{author}{\bibfnamefont{M.~S.} \bibnamefont{Frank}},
  \bibinfo{author}{\bibfnamefont{T.-H.} \bibnamefont{Lee}},
  \bibinfo{author}{\bibfnamefont{G.}~\bibnamefont{Bhattacharyya}},
  \bibinfo{author}{\bibfnamefont{P.~K.~H.} \bibnamefont{Tsang}},
  \bibinfo{author}{\bibfnamefont{V.~L.} \bibnamefont{Quito}},
  \bibinfo{author}{\bibfnamefont{V.}~\bibnamefont{Dobrosavljevi\ifmmode~\acute{c}\else
  \'{c}\fi{}}}, \bibinfo{author}{\bibfnamefont{O.}~\bibnamefont{Christiansen}},
  \bibnamefont{and} \bibinfo{author}{\bibfnamefont{N.}~\bibnamefont{Lanat\`a}},
  \bibinfo{journal}{Phys. Rev. B} \textbf{\bibinfo{volume}{104}},
  \bibinfo{pages}{L081103} (\bibinfo{year}{2021}),
  \urlprefix\url{https://link.aps.org/doi/10.1103/PhysRevB.104.L081103}.

\bibitem[{\citenamefont{Lanat\`a}(2022)}]{gRISB_2022}
\bibinfo{author}{\bibfnamefont{N.}~\bibnamefont{Lanat\`a}},
  \bibinfo{journal}{Phys. Rev. B} \textbf{\bibinfo{volume}{105}},
  \bibinfo{pages}{045111} (\bibinfo{year}{2022}),
  \urlprefix\url{https://link.aps.org/doi/10.1103/PhysRevB.105.045111}.

\bibitem[{\citenamefont{Lee et~al.}(2023{\natexlab{a}})\citenamefont{Lee,
  Lanat\`a, and Kotliar}}]{Lee_gRISB_2023_1}
\bibinfo{author}{\bibfnamefont{T.-H.} \bibnamefont{Lee}},
  \bibinfo{author}{\bibfnamefont{N.}~\bibnamefont{Lanat\`a}}, \bibnamefont{and}
  \bibinfo{author}{\bibfnamefont{G.}~\bibnamefont{Kotliar}},
  \bibinfo{journal}{Phys. Rev. B} \textbf{\bibinfo{volume}{107}},
  \bibinfo{pages}{L121104} (\bibinfo{year}{2023}{\natexlab{a}}),
  \urlprefix\url{https://link.aps.org/doi/10.1103/PhysRevB.107.L121104}.

\bibitem[{\citenamefont{Lee et~al.}(2023{\natexlab{b}})\citenamefont{Lee,
  Melnick, Adler, Lanat\`a, and Kotliar}}]{Lee_gRISB_2023_2}
\bibinfo{author}{\bibfnamefont{T.-H.} \bibnamefont{Lee}},
  \bibinfo{author}{\bibfnamefont{C.}~\bibnamefont{Melnick}},
  \bibinfo{author}{\bibfnamefont{R.}~\bibnamefont{Adler}},
  \bibinfo{author}{\bibfnamefont{N.}~\bibnamefont{Lanat\`a}}, \bibnamefont{and}
  \bibinfo{author}{\bibfnamefont{G.}~\bibnamefont{Kotliar}},
  \bibinfo{journal}{Phys. Rev. B} \textbf{\bibinfo{volume}{108}},
  \bibinfo{pages}{245147} (\bibinfo{year}{2023}{\natexlab{b}}),
  \urlprefix\url{https://link.aps.org/doi/10.1103/PhysRevB.108.245147}.

\bibitem[{\citenamefont{Giuli et~al.}(2026)\citenamefont{Giuli, Lee, Yao,
  Kotliar, Ruckenstein, Gingras, and Lanat\`a}}]{giuli2026}
\bibinfo{author}{\bibfnamefont{S.}~\bibnamefont{Giuli}},
  \bibinfo{author}{\bibfnamefont{T.-H.} \bibnamefont{Lee}},
  \bibinfo{author}{\bibfnamefont{Y.-X.} \bibnamefont{Yao}},
  \bibinfo{author}{\bibfnamefont{G.}~\bibnamefont{Kotliar}},
  \bibinfo{author}{\bibfnamefont{A.~E.} \bibnamefont{Ruckenstein}},
  \bibinfo{author}{\bibfnamefont{O.}~\bibnamefont{Gingras}}, \bibnamefont{and}
  \bibinfo{author}{\bibfnamefont{N.}~\bibnamefont{Lanat\`a}},
  \emph{\bibinfo{title}{Unifying variational and dynamical quantum embedding:
  From ghost {G}utzwiller approximation to dynamical mean-field theory}}
  (\bibinfo{year}{2026}), \eprint{2603.20559},
  \urlprefix\url{https://arxiv.org/abs/2603.20559}.

\bibitem[{\citenamefont{Mejuto-Zaera and
  Fabrizio}(2023)}]{Carlos_multiorb_2023}
\bibinfo{author}{\bibfnamefont{C.}~\bibnamefont{Mejuto-Zaera}}
  \bibnamefont{and} \bibinfo{author}{\bibfnamefont{M.}~\bibnamefont{Fabrizio}},
  \bibinfo{journal}{Phys. Rev. B} \textbf{\bibinfo{volume}{107}},
  \bibinfo{pages}{235150} (\bibinfo{year}{2023}),
  \urlprefix\url{https://link.aps.org/doi/10.1103/PhysRevB.107.235150}.

\bibitem[{\citenamefont{Lee et~al.}(2024)\citenamefont{Lee, Melnick, Adler,
  Sun, Yao, Lanat\`a, and Kotliar}}]{Lee_gRISB_2024}
\bibinfo{author}{\bibfnamefont{T.-H.} \bibnamefont{Lee}},
  \bibinfo{author}{\bibfnamefont{C.}~\bibnamefont{Melnick}},
  \bibinfo{author}{\bibfnamefont{R.}~\bibnamefont{Adler}},
  \bibinfo{author}{\bibfnamefont{X.}~\bibnamefont{Sun}},
  \bibinfo{author}{\bibfnamefont{Y.}~\bibnamefont{Yao}},
  \bibinfo{author}{\bibfnamefont{N.}~\bibnamefont{Lanat\`a}}, \bibnamefont{and}
  \bibinfo{author}{\bibfnamefont{G.}~\bibnamefont{Kotliar}},
  \bibinfo{journal}{Phys. Rev. B} \textbf{\bibinfo{volume}{110}},
  \bibinfo{pages}{115126} (\bibinfo{year}{2024}),
  \urlprefix\url{https://link.aps.org/doi/10.1103/PhysRevB.110.115126}.

\bibitem[{sup()}]{supplement}
\bibinfo{note}{See Supplemental Material at [URL will be inserted by publisher]
  for details of the derivation, additional data, and numerical
  implementation.}

\bibitem[{\citenamefont{Schir\'o and Fabrizio}(2010)}]{Schiro2010}
\bibinfo{author}{\bibfnamefont{M.}~\bibnamefont{Schir\'o}} \bibnamefont{and}
  \bibinfo{author}{\bibfnamefont{M.}~\bibnamefont{Fabrizio}},
  \bibinfo{journal}{Phys. Rev. Lett.} \textbf{\bibinfo{volume}{105}},
  \bibinfo{pages}{076401} (\bibinfo{year}{2010}).

\bibitem[{\citenamefont{Fabrizio}(2013)}]{Fabrizio2013}
\bibinfo{author}{\bibfnamefont{M.}~\bibnamefont{Fabrizio}}, in
  \emph{\bibinfo{booktitle}{New Materials for Thermoelectric Applications:
  Theory and Experiment}} (\bibinfo{publisher}{Springer},
  \bibinfo{address}{Dordrecht}, \bibinfo{year}{2013}), pp.
  \bibinfo{pages}{247--273}, \bibinfo{note}{arXiv:1204.2175}.

\bibitem[{\citenamefont{Fabrizio}(2017)}]{Fabrizio2017}
\bibinfo{author}{\bibfnamefont{M.}~\bibnamefont{Fabrizio}},
  \bibinfo{journal}{Phys. Rev. B} \textbf{\bibinfo{volume}{95}},
  \bibinfo{pages}{075156} (\bibinfo{year}{2017}),
  \urlprefix\url{https://link.aps.org/doi/10.1103/PhysRevB.95.075156}.

\bibitem[{\citenamefont{Guerci et~al.}(2023)\citenamefont{Guerci, Capone, and
  Lanat\`a}}]{Guerci2023}
\bibinfo{author}{\bibfnamefont{D.}~\bibnamefont{Guerci}},
  \bibinfo{author}{\bibfnamefont{M.}~\bibnamefont{Capone}}, \bibnamefont{and}
  \bibinfo{author}{\bibfnamefont{N.}~\bibnamefont{Lanat\`a}},
  \bibinfo{journal}{Phys. Rev. Research} \textbf{\bibinfo{volume}{5}},
  \bibinfo{pages}{L032023} (\bibinfo{year}{2023}).

\bibitem[{\citenamefont{Arrigoni and Strinati}(1995)}]{Arrigoni1995}
\bibinfo{author}{\bibfnamefont{E.}~\bibnamefont{Arrigoni}} \bibnamefont{and}
  \bibinfo{author}{\bibfnamefont{G.~C.} \bibnamefont{Strinati}},
  \bibinfo{journal}{Phys. Rev. B} \textbf{\bibinfo{volume}{52}},
  \bibinfo{pages}{2428} (\bibinfo{year}{1995}).

\bibitem[{\citenamefont{Riegler et~al.}(2026)\citenamefont{Riegler, Seufert,
  Thomale, and W\"olfle}}]{Riegler2026}
\bibinfo{author}{\bibfnamefont{D.}~\bibnamefont{Riegler}},
  \bibinfo{author}{\bibfnamefont{J.}~\bibnamefont{Seufert}},
  \bibinfo{author}{\bibfnamefont{R.}~\bibnamefont{Thomale}}, \bibnamefont{and}
  \bibinfo{author}{\bibfnamefont{P.}~\bibnamefont{W\"olfle}},
  \bibinfo{journal}{Phys. Rev. B} \textbf{\bibinfo{volume}{113}},
  \bibinfo{pages}{155131} (\bibinfo{year}{2026}).

\bibitem[{\citenamefont{Riegler et~al.}(2020)\citenamefont{Riegler, Klett,
  Neupert, Thomale, and W\"olfle}}]{Riegler2020}
\bibinfo{author}{\bibfnamefont{D.}~\bibnamefont{Riegler}},
  \bibinfo{author}{\bibfnamefont{M.}~\bibnamefont{Klett}},
  \bibinfo{author}{\bibfnamefont{T.}~\bibnamefont{Neupert}},
  \bibinfo{author}{\bibfnamefont{R.}~\bibnamefont{Thomale}}, \bibnamefont{and}
  \bibinfo{author}{\bibfnamefont{P.}~\bibnamefont{W\"olfle}},
  \bibinfo{journal}{Phys. Rev. B} \textbf{\bibinfo{volume}{101}},
  \bibinfo{pages}{235137} (\bibinfo{year}{2020}),
  \urlprefix\url{https://link.aps.org/doi/10.1103/PhysRevB.101.235137}.

\end{thebibliography}


\begin{thebibliography}{11}
\expandafter\ifx\csname natexlab\endcsname\relax\def\natexlab#1{#1}\fi
\expandafter\ifx\csname bibnamefont\endcsname\relax
  \def\bibnamefont#1{#1}\fi
\expandafter\ifx\csname bibfnamefont\endcsname\relax
  \def\bibfnamefont#1{#1}\fi
\expandafter\ifx\csname citenamefont\endcsname\relax
  \def\citenamefont#1{#1}\fi
\expandafter\ifx\csname url\endcsname\relax
  \def\url#1{\texttt{#1}}\fi
\expandafter\ifx\csname urlprefix\endcsname\relax\def\urlprefix{URL }\fi
\providecommand{\bibinfo}[2]{#2}
\providecommand{\eprint}[2][]{\url{#2}}

\bibitem[{\citenamefont{Giuli et~al.}(2026{\natexlab{a}})\citenamefont{Giuli,
  Lee, Yao, Kotliar, Ruckenstein, Gingras, and Lanat\`a}}]{giuli2026}
\bibinfo{author}{\bibfnamefont{S.}~\bibnamefont{Giuli}},
  \bibinfo{author}{\bibfnamefont{T.-H.} \bibnamefont{Lee}},
  \bibinfo{author}{\bibfnamefont{Y.-X.} \bibnamefont{Yao}},
  \bibinfo{author}{\bibfnamefont{G.}~\bibnamefont{Kotliar}},
  \bibinfo{author}{\bibfnamefont{A.~E.} \bibnamefont{Ruckenstein}},
  \bibinfo{author}{\bibfnamefont{O.}~\bibnamefont{Gingras}}, \bibnamefont{and}
  \bibinfo{author}{\bibfnamefont{N.}~\bibnamefont{Lanat\`a}},
  \emph{\bibinfo{title}{Unifying variational and dynamical quantum embedding:
  From ghost {G}utzwiller approximation to dynamical mean-field theory}}
  (\bibinfo{year}{2026}{\natexlab{a}}), \eprint{2603.20559},
  \urlprefix\url{https://arxiv.org/abs/2603.20559}.

\bibitem[{\citenamefont{Riegler et~al.}(2020)\citenamefont{Riegler, Klett,
  Neupert, Thomale, and W\"olfle}}]{Riegler2020}
\bibinfo{author}{\bibfnamefont{D.}~\bibnamefont{Riegler}},
  \bibinfo{author}{\bibfnamefont{M.}~\bibnamefont{Klett}},
  \bibinfo{author}{\bibfnamefont{T.}~\bibnamefont{Neupert}},
  \bibinfo{author}{\bibfnamefont{R.}~\bibnamefont{Thomale}}, \bibnamefont{and}
  \bibinfo{author}{\bibfnamefont{P.}~\bibnamefont{W\"olfle}},
  \bibinfo{journal}{Phys. Rev. B} \textbf{\bibinfo{volume}{101}},
  \bibinfo{pages}{235137} (\bibinfo{year}{2020}),
  \urlprefix\url{https://link.aps.org/doi/10.1103/PhysRevB.101.235137}.

\bibitem[{\citenamefont{Yao et~al.}(2021)\citenamefont{Yao, Zhang, Wang, Ho,
  and Orth}}]{Yao2021}
\bibinfo{author}{\bibfnamefont{Y.}~\bibnamefont{Yao}},
  \bibinfo{author}{\bibfnamefont{F.}~\bibnamefont{Zhang}},
  \bibinfo{author}{\bibfnamefont{C.-Z.} \bibnamefont{Wang}},
  \bibinfo{author}{\bibfnamefont{K.-M.} \bibnamefont{Ho}}, \bibnamefont{and}
  \bibinfo{author}{\bibfnamefont{P.~P.} \bibnamefont{Orth}},
  \bibinfo{journal}{Phys. Rev. Research} \textbf{\bibinfo{volume}{3}},
  \bibinfo{pages}{013184} (\bibinfo{year}{2021}).

\bibitem[{\citenamefont{Giuli et~al.}(2026{\natexlab{b}})\citenamefont{Giuli,
  Hasan, Kloss, Frank, Lee, Gingras, Yao, and Lanat\`a}}]{PCA2026}
\bibinfo{author}{\bibfnamefont{S.}~\bibnamefont{Giuli}},
  \bibinfo{author}{\bibfnamefont{H.}~\bibnamefont{Hasan}},
  \bibinfo{author}{\bibfnamefont{B.}~\bibnamefont{Kloss}},
  \bibinfo{author}{\bibfnamefont{M.~S.} \bibnamefont{Frank}},
  \bibinfo{author}{\bibfnamefont{T.-H.} \bibnamefont{Lee}},
  \bibinfo{author}{\bibfnamefont{O.}~\bibnamefont{Gingras}},
  \bibinfo{author}{\bibfnamefont{Y.-X.} \bibnamefont{Yao}}, \bibnamefont{and}
  \bibinfo{author}{\bibfnamefont{N.}~\bibnamefont{Lanat\`a}},
  \bibinfo{journal}{PRX Intelligence} \textbf{\bibinfo{volume}{1}},
  \bibinfo{pages}{013001} (\bibinfo{year}{2026}{\natexlab{b}}),
  \urlprefix\url{https://link.aps.org/doi/10.1103/phk3-pvhm}.

\bibitem[{\citenamefont{Besserve and Ayral}(2022)}]{Besserve2022}
\bibinfo{author}{\bibfnamefont{P.}~\bibnamefont{Besserve}} \bibnamefont{and}
  \bibinfo{author}{\bibfnamefont{T.}~\bibnamefont{Ayral}},
  \bibinfo{journal}{Phys. Rev. B} \textbf{\bibinfo{volume}{105}},
  \bibinfo{pages}{115108} (\bibinfo{year}{2022}).

\bibitem[{\citenamefont{Chen et~al.}(2026)\citenamefont{Chen, Khindanov,
  Salazar, Barona, Harrabi, Zhang, Wang, Iadecola, Lanat{\`a}, and
  Yao}}]{Chen2026}
\bibinfo{author}{\bibfnamefont{I.-C.} \bibnamefont{Chen}},
  \bibinfo{author}{\bibfnamefont{A.}~\bibnamefont{Khindanov}},
  \bibinfo{author}{\bibfnamefont{C.~M.} \bibnamefont{Salazar}},
  \bibinfo{author}{\bibfnamefont{H.~M.} \bibnamefont{Barona}},
  \bibinfo{author}{\bibfnamefont{G.}~\bibnamefont{Harrabi}},
  \bibinfo{author}{\bibfnamefont{F.}~\bibnamefont{Zhang}},
  \bibinfo{author}{\bibfnamefont{C.-Z.} \bibnamefont{Wang}},
  \bibinfo{author}{\bibfnamefont{T.}~\bibnamefont{Iadecola}},
  \bibinfo{author}{\bibfnamefont{N.}~\bibnamefont{Lanat{\`a}}},
  \bibnamefont{and} \bibinfo{author}{\bibfnamefont{Y.-X.} \bibnamefont{Yao}},
  \bibinfo{journal}{npj Quantum Information} \textbf{\bibinfo{volume}{12}},
  \bibinfo{pages}{102} (\bibinfo{year}{2026}).

\bibitem[{\citenamefont{Adler et~al.}(2024)\citenamefont{Adler, Melnick, and
  Kotliar}}]{adler_portobello_2024}
\bibinfo{author}{\bibfnamefont{R.}~\bibnamefont{Adler}},
  \bibinfo{author}{\bibfnamefont{C.}~\bibnamefont{Melnick}}, \bibnamefont{and}
  \bibinfo{author}{\bibfnamefont{G.}~\bibnamefont{Kotliar}},
  \bibinfo{journal}{Computer Physics Communications}
  \textbf{\bibinfo{volume}{294}}, \bibinfo{pages}{108907}
  (\bibinfo{year}{2024}), ISSN \bibinfo{issn}{0010-4655},
  \urlprefix\url{https://www.sciencedirect.com/science/article/pii/S0010465523002527}.

\bibitem[{\citenamefont{Melnick et~al.}(2021)\citenamefont{Melnick, Sémon, Yu,
  D'Imperio, Tremblay, and Kotliar}}]{melnick_accelerated_2021}
\bibinfo{author}{\bibfnamefont{C.}~\bibnamefont{Melnick}},
  \bibinfo{author}{\bibfnamefont{P.}~\bibnamefont{Sémon}},
  \bibinfo{author}{\bibfnamefont{K.}~\bibnamefont{Yu}},
  \bibinfo{author}{\bibfnamefont{N.}~\bibnamefont{D'Imperio}},
  \bibinfo{author}{\bibfnamefont{A.-M.} \bibnamefont{Tremblay}},
  \bibnamefont{and} \bibinfo{author}{\bibfnamefont{G.}~\bibnamefont{Kotliar}},
  \bibinfo{journal}{Computer Physics Communications}
  \textbf{\bibinfo{volume}{267}}, \bibinfo{pages}{108075}
  (\bibinfo{year}{2021}), ISSN \bibinfo{issn}{0010-4655},
  \urlprefix\url{https://www.sciencedirect.com/science/article/pii/S0010465521001879}.

\bibitem[{\citenamefont{Gunacker et~al.}(2015)\citenamefont{Gunacker,
  Wallerberger, Gull, Hausoel, Sangiovanni, and
  Held}}]{gunacker_continuous-time_2015}
\bibinfo{author}{\bibfnamefont{P.}~\bibnamefont{Gunacker}},
  \bibinfo{author}{\bibfnamefont{M.}~\bibnamefont{Wallerberger}},
  \bibinfo{author}{\bibfnamefont{E.}~\bibnamefont{Gull}},
  \bibinfo{author}{\bibfnamefont{A.}~\bibnamefont{Hausoel}},
  \bibinfo{author}{\bibfnamefont{G.}~\bibnamefont{Sangiovanni}},
  \bibnamefont{and} \bibinfo{author}{\bibfnamefont{K.}~\bibnamefont{Held}},
  \bibinfo{journal}{Phys. Rev. B} \textbf{\bibinfo{volume}{92}},
  \bibinfo{pages}{155102} (\bibinfo{year}{2015}), ISSN
  \bibinfo{issn}{1550235X},
  \urlprefix\url{https://link.aps.org/doi/10.1103/PhysRevB.92.155102}.

\bibitem[{\citenamefont{Gunacker et~al.}(2016)\citenamefont{Gunacker,
  Wallerberger, Ribic, Hausoel, Sangiovanni, and
  Held}}]{gunacker_worm-improved_2016}
\bibinfo{author}{\bibfnamefont{P.}~\bibnamefont{Gunacker}},
  \bibinfo{author}{\bibfnamefont{M.}~\bibnamefont{Wallerberger}},
  \bibinfo{author}{\bibfnamefont{T.}~\bibnamefont{Ribic}},
  \bibinfo{author}{\bibfnamefont{A.}~\bibnamefont{Hausoel}},
  \bibinfo{author}{\bibfnamefont{G.}~\bibnamefont{Sangiovanni}},
  \bibnamefont{and} \bibinfo{author}{\bibfnamefont{K.}~\bibnamefont{Held}},
  \bibinfo{journal}{Phys. Rev. B} \textbf{\bibinfo{volume}{94}},
  \bibinfo{pages}{125153} (\bibinfo{year}{2016}), ISSN
  \bibinfo{issn}{24699969},
  \urlprefix\url{https://link.aps.org/doi/10.1103/PhysRevB.94.125153}.

\bibitem[{\citenamefont{Kaufmann et~al.}(2017)\citenamefont{Kaufmann, Gunacker,
  and Held}}]{kaufmann_continuous-time_2017}
\bibinfo{author}{\bibfnamefont{J.}~\bibnamefont{Kaufmann}},
  \bibinfo{author}{\bibfnamefont{P.}~\bibnamefont{Gunacker}}, \bibnamefont{and}
  \bibinfo{author}{\bibfnamefont{K.}~\bibnamefont{Held}},
  \bibinfo{journal}{Phys. Rev. B} \textbf{\bibinfo{volume}{96}},
  \bibinfo{pages}{35114} (\bibinfo{year}{2017}), ISSN \bibinfo{issn}{24699969},
  \urlprefix\url{https://link.aps.org/doi/10.1103/PhysRevB.96.035114}.

\end{thebibliography}

\end{document}


\title{Supplemental Material for ``Ghost rotationally invariant slave-boson approach to response functions''}

\author{Tsung-Han Lee}
\affiliation{Department of Physics, National Chung Cheng University, Chiayi 62102, Taiwan}
\affiliation{Physics Division, National Center for Theoretical Sciences, Taipei 10617, Taiwan}

\author{Corey Melnick}
\affiliation{Physics and Astronomy Department, Rutgers University, Piscataway, New Jersey 08854, USA}
\affiliation{Condensed Matter Physics and Materials Science Department, Brookhaven National Laboratory, Upton, New York 11973, USA}

\author{Nicola Lanat\`a}
\affiliation{School of Physics and Astronomy, Rochester Institute of Technology, 84 Lomb Memorial Drive, Rochester, New York 14623, USA}
\affiliation{Center for Computational Quantum Physics, Flatiron Institute, New York, New York 10010, USA}

\author{Gabriel Kotliar}
\affiliation{Physics and Astronomy Department, Rutgers University, Piscataway, New Jersey 08854, USA}
\affiliation{Condensed Matter Physics and Materials Science Department, Brookhaven National Laboratory, Upton, New York 11973, USA}

\maketitle

\tableofcontents

\bigskip
Equation numbers (1)--(17) and Figs.~1--4 refer to the main text.

\section{The gRISB Lagrangian at finite temperature}
\label{sec:lagrangian}

The response theory requires the gRISB Lagrangian, Eq.~(2), to be written from the
outset for \emph{spatially inhomogeneous} and \emph{imaginary-time-dependent} field
configurations at \emph{finite temperature}: the slave-boson matrices are site and
imaginary-time dependent, $R_i(\tau),\lambda_i(\tau),\mathcal D_i(\tau),\lambda^c_i(\tau)$
[equivalently, they carry momentum and Matsubara frequency,
$R_{\bm q,\Om_n}=(\beta N)^{-1}\sum_i\int_0^\beta d\tau\,
e^{-i\bm q\cdot\bm R_i+i\Om_n\tau}R_i(\tau)$, etc.], and every term is an
imaginary-time-ordered thermal trace rather than a ground-state expectation value.
Only with the full
$(\bm k_1,\bm k_2)$ structure of $\mathcal L_{\mathrm{qp}}$, Eqs.~(4)--(6), the
imaginary-time ordering $\mathcal T_\tau$, and the
site resolution of $\mathcal L_{\mathrm{emb}}$ and $\mathcal L_{\mathrm{mix}}$,
Eqs.~(7)--(10), can a source of finite momentum $\bm q$ and frequency $i\Om_n$ be
attached to either sector, so that $\chi(\bm q,i\Om_n)$ is obtained by twice
differentiating $\mathcal L$.
For the response calculation, these time-ordered traces are used only through their
imaginary-time linked-cluster expansion to quadratic order about the static saddle, as
derived in Sec.~\ref{sec:flucblocks}.

At the spatially homogeneous static saddle point, $R_{\bm q,\Om_n}=R\,\delta_{\bm q,\bm 0}\delta_{\Om_n,0}$ and
$\lambda_{\bm q,\Om_n}=\lambda\,\delta_{\bm q,\bm 0}\delta_{\Om_n,0}$, the time
ordering trivializes, $\int_0^\beta d\tau\to\beta$, the momentum matrix of Eq.~(6)
becomes diagonal,
$[\underline H^{\mathrm{qp}}]_{\bm k_1\bm k_2}=\delta_{\bm k_1\bm k_2}\Hqp_{\bm k}$
with
\begin{equation}
\Hqp_{\bm k} = R\,\eps_{\bm k}\,R^\dg + \lambda ,
\label{eq:Hqphom}
\end{equation}
and $\mathcal L_{\mathrm{qp}}$ reduces to the quasiparticle grand potential
$-\tfrac TN\sum_{\bm k,\omega_n}\Tr\log[-i\omega_n+\Hqp_{\bm k}]\,e^{i\omega_n0^+}$,
a free-fermion determinant; the interacting $\mathcal L_{\mathrm{emb}}$ is instead
evaluated by exact diagonalization.

The mixing part $\mathcal L_{\mathrm{mix}}$, Eqs.~(9)--(10), deserves two remarks.
First, it is the free energy of
the \emph{non-interacting} auxiliary Hamiltonian $\hat H^{\mathrm{emb}}_{0,i}$~\cite{giuli2026},
a free-fermion object, which is what allows its fluctuation Hessian to be continued
analytically to finite frequency in Sec.~\ref{sec:flucblocks}. Second, the bath term
of Eq.~(10) is \emph{anti-normal ordered}, $\lambda^c\,b\,b^\dg$: the constant $\tr\lambda^c_i$ that it generates is carried inside the thermal trace. %
Evaluating the trace explicitly,
\begin{equation}
\mathcal L_{\mathrm{mix}}
 = \frac1N\sum_i\Big[\,\tr\lambda^c_i-\frac{1}{\beta}\sum_m \ln\!\big(1+e^{-\beta\eps_{i,m}}\big)\Big],
\label{eq:Lmixeval}
\end{equation}
where $\eps_{i,m}$ are the eigenvalues of the normal-ordered single-particle matrix
in the combined quasiparticle--bath $(f,b)$ space,
\begin{equation}
H^{\mathrm{emb}}_{0,i} = \begin{pmatrix} \lambda_i & R_i \mathcal D_i^{T} \\[2pt]
 (R_i \mathcal D_i^{T})^\dg & -\lambda^c_i \end{pmatrix}.
\label{eq:buildH}
\end{equation}

The complex variational fields are parameterized in a symmetry-adapted
single-particle basis $\{h_s, h_s^r\}$,
\begin{equation}
R = \sum_s (r'_s + i\,r''_s)\,h_s^r,\quad
\lambda = \sum_s l_s\,h_s,\quad
\mathcal D = \sum_s (D'_s + i\,D''_s)\,h_s^r,\quad
\lambda^c = \sum_s l^c_s\,h_s ,
\label{eq:basis}
\end{equation}
so that the fluctuation vector is
\begin{align}
\bm x &\equiv (R',R'',\lambda,D',D'',\lambda^c) \nonumber\\
&\equiv\big(r'_0,\ldots,r'_{N_s^r-1};
r''_0,\ldots,r''_{N_s^r-1};
l_0,\ldots,l_{N_s-1};\nonumber\\
&\hspace{3.25em}D'_0,\ldots,D'_{N_s^r-1};
D''_0,\ldots,D''_{N_s^r-1};
l^c_0,\ldots,l^c_{N_s-1}\big),
\end{align}
with dimension
$N_M = 4N^r_s+2N_s$, where $N^r_s$ ($N_s$) is the number of $R/\mathcal D$-type
($\lambda$-type) basis elements. For the one-band model with two additional ghost orbitals, there are $N_a=6$ quasiparticle spin-orbitals; we then need $N^r_s=3$ $R$-type and $N_s=6$ $\lambda$-type basis elements, so $N_M=24$ in total per
charge and spin channel, where $\{h_s, h_s^r\}$ is the most general basis compatible with the symmetries of the paramagnetic saddle (real matrices, diagonal in the spin structure of each channel) spanning the $R$/$\mathcal{D}$ and $\lambda$/$\lambda^c$ matrices. Explicitly,
each basis element is the tensor product of
an orbital factor with a spin factor $\sigma^{0}/\sqrt2$ (charge channel) or
$\sigma^{z}/\sqrt2$ (spin channel). The orbital factors used in the production
code are
\begin{equation}
e_{1}=\begin{pmatrix}1\\0\\0\end{pmatrix},\qquad
e_{2}=\begin{pmatrix}0\\1\\0\end{pmatrix},\qquad
e_{3}=\begin{pmatrix}0\\0\\1\end{pmatrix}
\label{eq:basisexp}
\end{equation}
for the three $R/\mathcal D$-type elements, $h^{r}_{s}=e_{s}\otimes\sigma^{0,z}/\sqrt2$,
and
\begin{equation}
\begin{gathered}
T_{1}=\begin{pmatrix}1&0&0\\0&0&0\\0&0&0\end{pmatrix},\qquad
T_{2}=\begin{pmatrix}0&0&0\\0&1&0\\0&0&0\end{pmatrix},\qquad
T_{3}=\begin{pmatrix}0&0&0\\0&0&0\\0&0&1\end{pmatrix},\\[4pt]
T_{4}=\frac{1}{\sqrt2}\begin{pmatrix}0&1&0\\1&0&0\\0&0&0\end{pmatrix},\qquad
T_{5}=\frac{1}{\sqrt2}\begin{pmatrix}0&0&1\\0&0&0\\1&0&0\end{pmatrix},\qquad
T_{6}=\frac{1}{\sqrt2}\begin{pmatrix}0&0&0\\0&0&1\\0&1&0\end{pmatrix}
\end{gathered}
\label{eq:basisT}
\end{equation}
for the six $\lambda/\lambda^{c}$-type elements, $h_{s}=T_{s}\otimes\sigma^{0,z}/\sqrt2$,
normalized as $\Tr(h^{\dg}_{s}h_{s'})=\delta_{ss'}$. This real basis is the one
assumed throughout for the charge and longitudinal ($S^{z}$) spin responses
computed here; for the transverse $S^{x},S^{y}$ channels the corresponding
complex (spin-off-diagonal) basis must be applied.

\section{Saddle point}
\label{sec:saddle}

At the spatially homogeneous static saddle the imaginary-time ordering is trivial and every
site is equivalent, so the Lagrangian
becomes a function of the four matrices
$(R,\lambda,\mathcal D,\lambda^c)$ alone:
\begin{equation}
\begin{aligned}
\mathcal L_{\rm qp} &= -\frac{1}{\beta N}\sum_{\bm k}\Tr\ln\!\big(1+e^{-\beta\Hqp_{\bm k}}\big),
 \qquad \Hqp_{\bm k}=R\,\eps_{\bm k}\,R^\dg+\lambda,\\
\mathcal L_{\rm emb} &= -\frac{1}{\beta}\ln\Tr\,e^{-\beta\Hemb}, \\
\mathcal L_{\rm mix} &= -\frac{1}{\beta}\Tr\ln\!\big(1+e^{-\beta H^{\mathrm{emb}}_{0}}\big)+\tr\lambda^c,
 \quad
 H^{\mathrm{emb}}_{0}=\begin{pmatrix}\lambda & R\mathcal D^{T}\\[2pt](R\mathcal D^{T})^\dg & -\lambda^c\end{pmatrix},
\end{aligned}
\label{eq:Lhom}
\end{equation}
with $\Hemb$ the interacting embedding of Eq.~(8), including the local interaction and chemical-potential terms, and $H^{\mathrm{emb}}_{0}$ the non-interacting one of
Eq.~\eqref{eq:buildH}; the $+\tr\lambda^c$ is the constant from the anti-normal bath ordering
[Eq.~\eqref{eq:Lmixeval}]. The quasiparticle and mixing traces are single-particle
(free-fermion); the embedding trace is over the interacting Fock space.

The saddle-point equations are the stationarity conditions $\partial\mathcal L/\partial\bm x=0$,
with the matrix shorthand $\bm x=(R,\lambda,\mathcal D,\lambda^c)$; its independent
real coordinates are those of Eq.~\eqref{eq:basis}. The free-fermion derivatives are
\begin{align}
\partial_x\mathcal L_{\rm qp}
 &=\frac1N\sum_{\bm k}\Tr[\nF(\Hqp_{\bm k})\,\partial_x\Hqp_{\bm k}],
\label{eq:dLqp}\\
\partial_x\mathcal L_{\rm mix}
 &=\Tr[\nF(H^{\mathrm{emb}}_{0})\,\partial_x H^{\mathrm{emb}}_{0}]
+\partial_x\tr\lambda^c.
\label{eq:dLmix}
\end{align}
The interacting term is the Hellmann--Feynman average
$\partial_x\mathcal L_{\rm emb}=\langle\partial_x\Hemb\rangle_{\rm emb}$, with
$\nF(H)=(1+e^{\beta H})^{-1}$ and
$\langle A\rangle_{\rm emb}=\Tr[A\,e^{-\beta\Hemb}]/\Tr\,e^{-\beta\Hemb}$ the finite-temperature
thermal average over the embedding spectrum. Writing the non-interacting embedding one-body
density matrix, i.e., the transpose of the Fermi function,
$\langle\psi^\dg_i\psi_j\rangle=[\nF(H^{\mathrm{emb}}_{0})]_{ji}$, in its quasiparticle ($f$) and
bath ($b$) blocks,
\begin{equation}
\rho^0\equiv\nF\!\big(H^{\mathrm{emb}}_{0}\big)^{T}
 =\begin{pmatrix}\rho^0_{ff}&\rho^0_{fb}\\[2pt]\rho^{0\dg}_{fb}&\rho^0_{bb}\end{pmatrix},
\label{eq:rho0}
\end{equation}
the four conditions read, with quasiparticle/bath indices $a,b,c,d$, physical spin
index $\sigma$, each $\rho^0$ block an $N_a\times N_a$ matrix in $(a,b)$, and
$[\Hqp_{\bm k}]_{ab}=\sum_{\sigma}R_{a\sigma}[\eps_{\bm k}]_{\sigma}R^*_{b\sigma}+\lambda_{ab}$:
\begin{align}
\partial_{\lambda}\!:\ \
 [\rho^0_{ff}]_{ab} &= \frac1N\sum_{\bm k}[\nF(\Hqp_{\bm k})]_{ba},
\label{eq:saddle_lam}\\
\partial_{R}\!:\ \
 \sum_{c}[\rho^0_{fb}]_{ac}\,\mathcal D_{c\sigma}
 &= \frac1N\sum_{\bm k}\sum_{d}[\eps_{\bm k}]_{\sigma}\,R^*_{d\sigma}\,[\nF(\Hqp_{\bm k})]_{da},
\label{eq:saddle_R}\\
\partial_{\lambda^c}\!:\ \
 [\rho^0_{bb}]_{ab} &= \langle b^\dg_a b_b\rangle_{\rm emb},
\label{eq:saddle_lamc}\\
\partial_{\mathcal D}\!:\ \
 \sum_{c}R_{c\sigma}\,[\rho^0_{fb}]_{ca} &= \langle c^\dg_\sigma b_a\rangle_{\rm emb},
\label{eq:saddle_D}
\end{align}
where the transposed indices on $[\nF(\Hqp_{\bm k})]_{ba}$ and $[\rho^0]_{ab}=[\nF(H^{\mathrm{emb}}_{0})]_{ba}$
reflect that the density matrix is the transpose of the Fermi function.
The $\partial_\lambda,\partial_R$ conditions arise because $\lambda,R$ appear in both
$\mathcal L_{\rm qp}$ and $H^{\mathrm{emb}}_{0}$, and the $\partial_{\lambda^c},\partial_{\mathcal D}$
conditions because $\lambda^c,\mathcal D$ appear in both $\mathcal L_{\rm emb}$ and $H^{\mathrm{emb}}_{0}$
[the $+\tr\lambda^c$ term is what makes the bath blocks match in $\partial_{\lambda^c}$].
Equations~\eqref{eq:saddle_lam}--\eqref{eq:saddle_D} are solved self-consistently for
$(R,\lambda,\mathcal D,\lambda^c)$; the interacting embedding (two impurity and $N_a$ bath
spin-orbitals) is solved by exact diagonalization, with thermal averages over the spectrum.
With the converged $R,\lambda,\mathcal D,\lambda^c$, the quasiparticle and physical
propagators are
\begin{equation}
\underline G_{\bm k}(i\omega_n)=\big[i\omega_n - \Hqp_{\bm k}\big]^{-1},
\qquad
G_{\sigma}(\bm k,i\omega_n)
 = \sum_{ab}R^\dg_{\sigma a}\,\big[\underline G_{\bm k}(i\omega_n)\big]_{ab}\, R_{b\sigma}.
\label{eq:Gphys}
\end{equation}
The total spectral weight of $G$ is $R^\dg R$. In RISB it equals the quasiparticle
weight $Z$, whereas in gRISB it also contains the ghost (Hubbard-band) poles, so that
the low-energy $Z$ is smaller (e.g., $Z=0.45$ against $R^\dg R=0.944$ at $n=0.75$, $U=10$).
Saddle points used for fluctuation calculations are converged to $10^{-8}$ or better.

\section{Derivation of the response formula}
\label{sec:derivation}

The susceptibility formula, Eq.~(11), follows from coupling a source to the physical
channel operator and expanding the free energy to second order. We first set up the
general space- and time-dependent construction, reduce it to a single Fourier block, and
then extract the response in four steps.

\emph{General source and dynamical Lagrangian.} A source $\eta_i(\tau)$, independent at
every site $i$ and imaginary time $\tau$, couples to the local physical channel operator
in the embedding sector,
\begin{equation}
\Hemb_i(\tau) \;\to\; \Hemb_i(\tau) - \eta_i(\tau)\,\hat{\mathcal O}_{i},
\qquad
\hat{\mathcal O}_{i} = \sum_{\sigma\sigma'}\mathcal O_{\sigma\sigma'}\,
   c^\dg_{i\sigma} c_{i\sigma'}.
\label{eq:source}
\end{equation}
Inserting this source into the dynamical Lagrangian of Eqs.~(4), (7), (9), which is already a
functional of the spatial- and imaginary-time-dependent fields $\bm x_i(\tau)$ [Eqs.~(2) and (3)]
with all thermal traces imaginary-time ordered, gives
the source-shifted Lagrangian
\begin{equation}
\mathcal L[\bm x;\eta]
 = \mathcal L_{\mathrm{qp}}[\bm x]
 + \mathcal L_{\mathrm{emb}}[\bm x;\eta]
 - \mathcal L_{\mathrm{mix}}[\bm x],
\label{eq:Ldyn}
\end{equation}
\begin{equation}
\begin{aligned}
\mathcal L_{\mathrm{qp}}[\bm x] &=
 -\frac{1}{\beta N}\ln\Tr\,\mathcal T_\tau
  \exp\!\Big[-\!\int_0^{\beta}\! d\tau\,\hat H^{\mathrm{qp}}\big(\{R_i(\tau),\lambda_i(\tau)\}\big)\Big],\\
\mathcal L_{\mathrm{emb}}[\bm x;\eta] &=
 -\frac{1}{\beta N}\sum_i\ln\Tr\,\mathcal T_\tau
  \exp\!\Big[-\!\int_0^{\beta}\! d\tau\,\big(\Hemb_i(\tau)-\eta_i(\tau)\,\hat{\mathcal O}_i\big)\Big],\\
\mathcal L_{\mathrm{mix}}[\bm x] &=
 -\frac{1}{\beta N}\sum_i\ln\Tr\,\mathcal T_\tau
  \exp\!\Big[-\!\int_0^{\beta}\! d\tau\,\hat H^{\mathrm{emb}}_{0,i}(\tau)\Big],
\end{aligned}
\label{eq:Lparts}
\end{equation}
with $\Hemb_i$/$\hat H^{\mathrm{emb}}_{0,i}$ the interacting/non-interacting embedding Hamiltonian of Eq.~(8)/(10),
carrying the $\tau$ dependence through $R_i(\tau)$, $\lambda_i(\tau)$, $\mathcal D_i(\tau)$,
$\lambda^c_i(\tau)$.
The gauge-covariant and gauge-invariant properties of the gRISB framework can only be enforced at all frequencies if the three Lagrangian terms are treated as imaginary-time-dependent simultaneously.

\emph{Reduction to a single Fourier block.} Expanding about the homogeneous static
saddle,
\begin{equation}
x_{\mu,i}(\tau)=x^{(0)}_\mu+\sum_{\bm q,\Om_n}\delta x_{\mu,\bm q,\Om_n}\,
  e^{i\bm q\cdot\bm R_i-i\Om_n\tau},
\qquad
\eta_i(\tau)=\sum_{\bm q,\Om_n}\eta_{\bm q,\Om_n}\,e^{i\bm q\cdot\bm R_i-i\Om_n\tau},
\label{eq:fourier}
\end{equation}
with $\Om_n=2\pi n/\beta$ bosonic and inverse
$\eta_{\bm q,\Om_n}=(\beta N)^{-1}\sum_i\int_0^\beta d\tau\,\eta_i(\tau)\,
e^{-i\bm q\cdot\bm R_i+i\Om_n\tau}$ (and likewise for $\delta x$). A real source obeys
$\eta_{-\bm q,-\Om_n}=\eta^*_{\bm q,\Om_n}$, which is why the two sources attached below carry
opposite labels. By
space--time translation invariance of the saddle the quadratic action is
block-diagonal in $(\bm q,\Om_n)$: each block is governed by the fluctuation matrix
$\mathcal M(\bm q,i\Om_n)$ of Eq.~(12), and a source $\eta_{\bm q,\Om_n}$ induces only the
matching component $\delta x_{\mu,\bm q,\Om_n}$ at linear order. A single
momentum/frequency block therefore suffices, so Eq.~\eqref{eq:source} can be projected
onto that harmonic, $\eta_i(\tau)=\eta_{\bm q}(\tau)\,e^{i\bm q\cdot\bm R_i}$ with
$\eta_{\bm q}(\tau)=\eta_{\bm q,\Om_n}\,e^{-i\Om_n\tau}$, as used below.

\emph{(i) Generating functional.} Attach two sources of opposite momentum and
frequency, $\eta_1\equiv\eta_{-\bm q,-\Om_n}$ coupled to $\mathcal O_1$ and
$\eta_2\equiv\eta_{\bm q,\Om_n}$ coupled to $\mathcal O_2$. At fixed sources, the
fields are re-stationarized at $\bm x^{\star}(\eta)$ solving
$\partial\mathcal L/\partial x_{\mu,\bm q',\Om_{n'}}\big|_{x^{\star}(\eta)}=0$ for every
$(\mu,\bm q',\Om_{n'})$, and
\begin{equation}
F(\eta_1,\eta_2) = \mathcal L\big[x^{\star}(\eta);\eta\big],
\qquad
\chi_{\mathcal O_1\mathcal O_2}(\bm q,i\Om_n)
 = -\,\frac{d^2 F}
   {d\eta_{-\bm q,-\Om_n}\,d\eta_{\bm q,\Om_n}}\bigg|_{\eta=0}.
\label{eq:genfun}
\end{equation}
Here, the generating functional $F$ is the Lagrangian $\mathcal{L}$ restricted to the stationary path $\bm x^{\star}$, and only the induced $(\bm q,\Om_n)$ Fourier component enters, by the block-diagonality
established above. Also, $\bm x^{\star}=\bm x^{(0)}$, the homogeneous saddle point, when $\eta=0$.

\emph{(ii) First-order stationarity.} Differentiating $F$ once along the $\bm x^{\star}$ path, the saddle shift drops out
by stationarity,
\begin{equation}
\frac{dF}{d\eta_{\bm q,\Om_n}}
 = \frac{\partial\mathcal L}{\partial\eta_{\bm q,\Om_n}}
 + \sum_{\nu}\underbrace{\frac{\partial\mathcal L}{\partial x_{\nu,\bm q,\Om_n}}}
   _{=\,0\ \text{at}\ \bm x^{\star}}
   \frac{dx^{\star}_{\nu,\bm q,\Om_n}}{d\eta_{\bm q,\Om_n}}
 = \frac{\partial\mathcal L}{\partial\eta_{\bm q,\Om_n}},
\label{eq:envelope}
\end{equation}
so the response comes entirely from the second derivative, where the shift no longer
cancels.

\emph{(iii) Induced saddle shift.} Differentiating the stationarity condition
$\partial\mathcal L/\partial x_{\mu,-\bm q,-\Om_n}=0$ with respect to
$\eta_{\bm q,\Om_n}$ along the $\bm x^{\star}$ path, and recognizing the Hessian block
as the fluctuation matrix of Eq.~(12),
$[\mathcal M(\bm q,i\Om_n)]_{\mu\nu}
 =\partial^2\mathcal L/\partial x_{\mu,-\bm q,-\Om_n}\,\partial x_{\nu,\bm q,\Om_n}$, we obtain
\begin{equation}
\begin{aligned}
&\sum_\nu \big[\mathcal M(\bm q,i\Om_n)\big]_{\mu\nu}\,
 \frac{dx^{\star}_{\nu,\bm q,\Om_n}}{d\eta_{\bm q,\Om_n}}
 + \frac{\partial^2\mathcal L}
   {\partial x_{\mu,-\bm q,-\Om_n}\,\partial\eta_{\bm q,\Om_n}} = 0 ,\\[2pt]
&\frac{dx^{\star}_{\nu,\bm q,\Om_n}}{d\eta_{\bm q,\Om_n}}
 = -\sum_\mu\big[\mathcal M(\bm q,i\Om_n)\big]^{-1}_{\nu\mu}\,
   \frac{\partial^2\mathcal L}
   {\partial x_{\mu,-\bm q,-\Om_n}\,\partial\eta_{\bm q,\Om_n}} ,
\end{aligned}
\label{eq:saddleshift}
\end{equation}
which is the dynamical fluctuation of the coordinate $\delta x_{\nu,\bm q,\Om_n}$
induced at linear order by the source.

\emph{(iv) Second derivative.} Define the shorthand
\begin{equation}
\mathcal L_{ab}\equiv
\left.\frac{\partial^2\mathcal L}{\partial a\,\partial b}\right|_{\eta=0}.
\label{eq:Lab}
\end{equation}
Differentiating Eq.~\eqref{eq:envelope} once more and inserting Eq.~\eqref{eq:saddleshift} gives
\begin{equation}
\chi_{\mathcal O_1\mathcal O_2}(\bm q,i\Om_n)
 = -\frac{d^2F}{d\eta_1\, d\eta_2}\bigg|_{\eta=0}
 = \underbrace{-\,\mathcal L_{\eta_1\eta_2}}_{\displaystyle
   \chi^{\mathrm{emb}}_{\mathcal O_1\mathcal O_2}(i\Om_n)}
 + \sum_{\mu\nu} \underbrace{\mathcal L_{\eta_1 x_\mu}}_{\displaystyle
   \chi^{\mathrm{emb}}_{\mathcal O_1 x_\mu}(i\Om_n)}
   \big[\mathcal M(\bm q,i\Om_n)\big]^{-1}_{\mu\nu}
   \underbrace{\mathcal L_{x_\nu \eta_2}}_{\displaystyle
   \chi^{\mathrm{emb}}_{x_\nu\mathcal O_2}(i\Om_n)},
\label{eq:chain}
\end{equation}
which is Eq.~(11) of the main text for $\mathcal O_1=\mathcal O_2=\mathcal O$. Written out with the momentum--frequency labels,
$\eta_1=\eta_{-\bm q,-\Om_n}$ and $\eta_2=\eta_{\bm q,\Om_n}$, the bare susceptibility
is the Lehmann response of $\Hemb$,
$\chi^{\mathrm{emb}}_{\mathcal O_1\mathcal O_2}
 =-\partial^2\mathcal L/\partial\eta_{-\bm q,-\Om_n}\partial\eta_{\bm q,\Om_n}|_{\eta=0}$,
and the source vectors are the mixed derivatives
$\chi^{\mathrm{emb}}_{\mathcal O_1 x_\mu}(i\Om_n)
 =\partial^2\mathcal L/\partial\eta_{-\bm q,-\Om_n}\partial x_{\mu,\bm q,\Om_n}|_{\eta=0}$ and
$\chi^{\mathrm{emb}}_{x_\nu\mathcal O_2}(i\Om_n)
 =\partial^2\mathcal L/\partial x_{\nu,-\bm q,-\Om_n}\partial\eta_{\bm q,\Om_n}|_{\eta=0}$,
nonzero only on the embedding directions $(D',D'',\lambda^c)$. These response functions can be computed using the Lehmann representation shown later in Eq.~\eqref{eq:thermallehmann}.
The general four-index response tensor is
recovered by letting $\mathcal O_1,\mathcal O_2$ run over a basis of elementary
spin-orbital matrices; for the charge and spin channels of this work,
$\mathcal O_1=\mathcal O_2=\sigma^0/\sqrt2$ or $\sigma^z/\sqrt2$
(Sec.~\ref{sec:numerics}).

The same construction with the source attached instead to the quasiparticle sector
[Eq.~\eqref{eq:cqpasource}] defines the quasiparticle
(Green's-function--source) route underlying the conserving quasiparticle
approximation of the main text; it is carried out in Sec.~\ref{sec:cqpa}.

\section{The fluctuation matrix}
\label{sec:flucblocks}

The fluctuation matrix decomposes into the Hessians of the three Lagrangian terms,
\begin{equation}
\mathcal{M}_{\mu\nu}(\bm q,i\Om_n)
 = \Mqp_{\mu\nu}(\bm q,i\Om_n) + \Mmix_{\mu\nu}(i\Om_n) + \Memb_{\mu\nu}(i\Om_n).
\label{eq:Mdecomp}
\end{equation}
The quasiparticle block occupies the $(R',R'',\lambda)$ directions, the embedding
block the $(D',D'',\lambda^c)$ directions, and the mixing block couples them.

\emph{Second variation of a trace-log.} Every term of Eq.~\eqref{eq:Ldyn} has the form
$\tfrac{s}{\beta N}\ln\Tr\,\mathcal T_\tau e^{-S}$,
$S=\int_0^\beta\! d\tau\,\hat H(\bm x(\tau))$, with $s=-1$ for
$\mathcal L_{\mathrm{qp}},\mathcal L_{\mathrm{emb}}$ and $s=+1$ for the
mixing term $-\mathcal L_{\mathrm{mix}}$. Writing
$\hat H(\bm x^{(0)}{+}\delta\bm x)=\hat H^{(0)}+\sum_\mu\hat H'_\mu\,\delta x_\mu(\tau)
+\tfrac12\sum_{\mu\nu}\hat H''_{\mu\nu}\,\delta x_\mu(\tau)\delta x_\nu(\tau)$
[$\hat H'_\mu\equiv\partial\hat H/\partial x_\mu$ the single-vertex operator and
$\hat H''_{\mu\nu}\equiv\partial^2\hat H/\partial x_\mu\partial x_\nu$ the equal-time
contact, both at the homogeneous saddle $\bm x^{(0)}$], so that
$\delta S=\int_0^\beta\! d\tau\,\delta\hat H(\tau)$, the linked-cluster (cumulant) expansion
$\ln\Tr\,\mathcal T_\tau e^{-S}=\ln Z_0-\langle\delta S\rangle
+\tfrac12\langle\delta S^2\rangle_c$ ($\langle\cdot\rangle$ the thermal average, subscript
$c$ connected) has a linear piece that vanishes by the saddle condition of
Sec.~\ref{sec:saddle} [only its $\bm q{=}0,\Om_n{=}0$ component survives
$\int_0^\beta\! d\tau\sum_i$, and there it is the saddle gradient], and a quadratic piece
built from the equal-time contact $\langle\hat H''_{\mu\nu}\rangle$ and the two-time
connected correlator $\langle\mathcal T_\tau\,\hat H'_\mu(\tau)\,\hat H'_\nu(0)\rangle_c$.
Each fluctuation block of Eq.~\eqref{eq:Mdecomp} is its bosonic Matsubara transform,
\begin{equation}
\mathcal M^{p}_{\mu\nu}(\bm q,i\Om_n)
 = s\Big[\int_0^\beta\! d\tau\,e^{i\Om_n\tau}\,
   \langle\mathcal T_\tau\,\hat H'_\mu(\tau)\,\hat H'_\nu(0)\rangle_c
 - \langle\hat H''_{\mu\nu}\rangle\,\Big],\quad p\in\{\mathrm{qp, mix, emb}\},
\label{eq:secondvar}
\end{equation}
with $s$ fixing the relative signs of the three blocks; the contact
$\langle\hat H''_{\mu\nu}\rangle$ is the equal-time (diamagnetic) term and the correlator the
paramagnetic bubble. For the two \emph{free-fermion} blocks the bubble is a Wick loop, whose
fermionic Matsubara sum
$\int_0^\beta\! d\tau\,e^{i\Om_n\tau}\langle\mathcal T_\tau\,\hat H'_\mu(\tau)\,\hat H'_\nu(0)\rangle_c
=-\tfrac1\beta\sum_{\omega_n}\tr[\Lambda_\mu\,\underline G(i\omega_n{+}i\Om_n)\,\Lambda_\nu\,
\underline G(i\omega_n)]$
[$\underline G(i\omega_n)=(i\omega_n-\hat H^{(0)})^{-1}$ the sector's free-fermion propagator]
combines with $s$: the $s=-1$ quasiparticle block gives
$+[\,\text{bubble}+\text{contact}\,]$ [Eq.~\eqref{eq:Mqp}], and the $s=+1$ mixing block
$-[\,\text{bubble}+\text{contact}\,]$ [Eq.~\eqref{eq:mmixan}]. The \emph{interacting} embedding
block ($s=-1$) has no Wick loop, so $s$ acts on the correlator directly,
$\Memb_{\mu\nu}(i\Om_n)=-\int_0^\beta\! d\tau\,e^{i\Om_n\tau}\,
\langle\mathcal T_\tau\,\hat H'_\mu(\tau)\,\hat H'_\nu(0)\rangle_c$
[Eq.~\eqref{eq:Membchi}], evaluated from the exact-diagonalization Lehmann function below.
All partial derivatives are taken at the homogeneous saddle $\bm x^{(0)}$. Momentum enters
only through the quasiparticle dispersion: in $\Hqp_{\bm k}=R\eps_{\bm k}R^\dg+\lambda$ the
lattice $\bm k$ sits in $\eps_{\bm k}$, so the vertices
$\Lambda_\mu=\partial\underline H^{\mathrm{qp}}/\partial x_\mu$ carry $\bm k$ and the transfer
$\bm q$ pairs $\underline G_{\bm k}$ with $\underline G_{\bm k+\bm q}$, leaving
$\Mqp(\bm q,i\Om_n)$ the only $\bm q$-dependent block. The embedding Hamiltonians
$\Hemb$ and $H^{\mathrm{emb}}_{0}$ are single-site (impurity plus finite bath) with
no momentum label, so $\Memb$ and $\Mmix$ reduce to single-site eigenstate sums,
independent of $\bm k$ and $\bm q$ and depending on frequency alone. The three blocks are
evaluated next.

\emph{Quasiparticle block.} $\Mqp$ is a Lindhard response of $\Hqp_{\bm k}$, with a
diamagnetic (Fermi-factor) contact and a paramagnetic bubble,
\begin{equation}
\Mqp_{\mu\nu}(\bm q,i\Om_n) = \frac{1}{N}\sum_{\bm k}
 \tr\!\Big\{ \nF(\Hqp_{\bm k})\,
 \frac{\partial^2 \underline H^{\mathrm{qp}}}{\partial x_{\mu,-\bm q}\partial x_{\nu,\bm q}}
  + \frac{1}{\beta}\!\sum_{\omega_n}\underline{G}_{\bm k+\bm q}(i\omega_n{+}i\Om_n)\,\Lambda^{\bm k+\bm q,\bm q}_\nu\,
    \underline{G}_{\bm k}(i\omega_n)\,\Lambda^{\bm k,-\bm q}_\mu \Big\},
\label{eq:Mqp}
\end{equation}
with $\underline G_{\bm k}$ of Eq.~\eqref{eq:Gphys} and the vertices
\begin{equation}
\Lambda^{\bm k+\bm q,\bm q}_{r'_s} = h_s^r\,\eps_{\bm k}\,R^\dg + R\,\eps_{\bm k+\bm q}\,h_s^{r\dg},
\qquad
\Lambda^{\bm k+\bm q,\bm q}_{r''_s} = i\,h_s^r\,\eps_{\bm k}\,R^\dg - i\,R\,\eps_{\bm k+\bm q}\,h_s^{r\dg},
\qquad
\Lambda^{\bm k+\bm q,\bm q}_{l_s} = h_s ,
\label{eq:vertices}
\end{equation}
and $\Lambda^{\bm k,-\bm q}_\mu=[\Lambda^{\bm k+\bm q,\bm q}_\mu]^\dg$, which equals the main-text vertex $\Lambda_\mu(\bm k,\bm q)$.
The two terms are the contact and bubble of Eq.~\eqref{eq:secondvar}: the contact
$\langle\hat H''_{\mu\nu}\rangle$ is nonzero only for $R$-$R$ pairs, since
$\Hqp_{\bm k}=R\eps_{\bm k}R^\dg+\lambda$ is quadratic in $R$ but linear in $\lambda$
[explicitly,
$\partial^{2}\underline H^{\mathrm{qp}}/\partial x_{\mu,-\bm q}\partial x_{\nu,\bm q}
 = g_{\mu}\,\eps_{\bm k+\bm q}\,g_{\nu}^{\dg} + g_{\nu}\,\eps_{\bm k-\bm q}\,g_{\mu}^{\dg}$
with $g_{r'_s}=h^{r}_{s}$, $g_{r''_s}=i h^{r}_{s}$, $g_{l_s}=0$], and the vertices $\Lambda_\mu=\partial\underline H^{\mathrm{qp}}/\partial x_\mu$ follow
from the basis expansion of Eq.~\eqref{eq:basis} [$\partial R/\partial r'_s=h_s^r$,
$\partial R/\partial r''_s=i h_s^r$, $\partial\lambda/\partial l_s=h_s$]. $\Mqp$ satisfies
$\Mqp(\bm q,i\Om_n)^\dg = \Mqp(-\bm q,-i\Om_n)$.

\emph{Mixing block.} Because $\mathcal L_{\mathrm{mix}}$ is the free energy of
the non-interacting $H^{\mathrm{emb}}_{0}$ of Eq.~\eqref{eq:buildH}, the Hessian of the mixing term $-\mathcal L_{\mathrm{mix}}$
evaluates analytically to a free-fermion Lindhard bubble plus a diamagnetic contact,
\begin{equation}
\Mmix_{\mu\nu}(i\Om_n) = -\Big[\,B^{\mathrm{para}}_{\mu\nu}(i\Om_n)
 + B^{\mathrm{dia}}_{\mu\nu}\,\Big],
\label{eq:mmixan}
\end{equation}
\begin{equation}
B^{\mathrm{para}}_{\mu\nu}(i\Om_n) = \sum_{mm'}
 \frac{\nF(\eps_m)-\nF(\eps_{m'})}{i\Om_n+\eps_m-\eps_{m'}}\,
 [\Lambda^0_\mu]_{mm'}\,[\Lambda^0_\nu]_{m'm},
\qquad
B^{\mathrm{dia}}_{\mu\nu} = \tr\!\Big[\nF(H^{\mathrm{emb}}_{0})\,
 \frac{\partial^2 H^{\mathrm{emb}}_{0}}{\partial x_\mu\partial x_\nu}\Big],
\label{eq:bparabdia}
\end{equation}
where $\eps_m$ are the eigenvalues of $H^{\mathrm{emb}}_{0}$ and
$\nF(H^{\mathrm{emb}}_{0})=\sum_m \nF(\eps_m)|m\rangle\langle m|$ is its Fermi matrix,
$\Lambda^0_\mu=\partial H^{\mathrm{emb}}_{0}/\partial x_\mu$, and the degenerate
($\eps_{m'}\!\to\!\eps_m$) limit of the bubble kernel is $\nF'(\eps_m)$.
Equation~\eqref{eq:bparabdia} is the free-fermion
loop of Eq.~\eqref{eq:secondvar} in the eigenbasis of $H^{\mathrm{emb}}_{0}$: the fermionic
Matsubara sum
\begin{equation}
\frac1\beta\sum_{\omega_n}\frac{1}{i\omega_n-\eps_m}\,
 \frac{1}{i\omega_n+i\Om_n-\eps_{m'}}
 = \frac{\nF(\eps_m)-\nF(\eps_{m'})}{i\Om_n+\eps_m-\eps_{m'}}
\label{eq:matsum}
\end{equation}
contracts with the
vertices $[\Lambda^0_\mu]_{mm'}[\Lambda^0_\nu]_{m'm}$ to give $B^{\mathrm{para}}$, while
$-\langle\hat H''\rangle=-B^{\mathrm{dia}}$ is the contact; the $s=+1$ prefactor of
the mixing term $-\mathcal L_{\mathrm{mix}}$ supplies the overall minus. The contact $B^{\mathrm{dia}}$
is frequency
independent and is nonzero only for $(R,\mathcal D)$ cross pairs, because the
hybridization $R\mathcal D^{T}$ in Eq.~\eqref{eq:buildH} is bilinear.

\emph{Embedding block and source vectors.} These follow directly from the second variation of
the interacting $\mathcal L_{\mathrm{emb}}=-\tfrac1{\beta N}\sum_i\ln\Tr\,\mathcal T_\tau
\exp[-\!\int_0^\beta\! d\tau\,(\Hemb_i-\eta_i\hat{\mathcal O}_i)]$. Both the source and the field
fluctuations enter only through $\Hemb$,
$\Hemb\to\Hemb-\eta\,\hat{\mathcal O}+\sum_\mu(\partial\Hemb/\partial x_\mu)\,\delta x_\mu$, and
because $\Hemb$ is linear in $(\mathcal D,\lambda^c)$ the contact $\hat H''$ vanishes; the
$s=-1$ identity of Eq.~\eqref{eq:secondvar} then reduces every embedding object to a single connected
imaginary-time correlator of $\Hemb$,
\begin{equation}
\chi^{\mathrm{emb}}_{AB}(\tau)\equiv\langle\mathcal T_\tau\,A(\tau)\,B(0)\rangle_c ,
\qquad A,B\in\{\hat{\mathcal O},\ \partial\Hemb/\partial x_\mu\}\ \ [\mu\in(D',D'',\lambda^c)] .
\label{eq:embcorr}
\end{equation}
Because $\Hemb$ is strictly on-site this correlator is site diagonal
[$\delta^2\mathcal L_{\mathrm{emb}}/\delta\eta_i\,\delta\eta_j\propto\delta_{ij}$], so the two
opposite-momentum phases $e^{\pm i\bm q\cdot\bm R_i}$ collapse onto the same site and cancel:
the embedding response is $\bm q$ independent, and all momentum structure of
$\chi(\bm q,i\Om_n)$ resides in $\Mqp$ and the quasiparticle legs. Equation~\eqref{eq:embcorr}
is evaluated by exact diagonalization: inserting the eigenstates
$\Hemb|m\rangle=E_m|m\rangle$ into the connected correlator gives, for $0<\tau<\beta$,
\begin{equation}
\langle\mathcal T_\tau\,A(\tau)\,B(0)\rangle_c = \frac1{\mathcal Z}\sum_{m,n}
 p_m\,e^{\tau(E_m-E_n)}\,\langle m|A|n\rangle\langle n|B|m\rangle
 - \langle A\rangle\langle B\rangle ,
\label{eq:embtau}
\end{equation}
whose diagonal ($m{=}n$) terms combine with $-\langle A\rangle\langle B\rangle$ into a
$\tau$-independent elastic constant. The bosonic Matsubara transform of the
$m\ne n$ terms,
$\int_0^\beta\! d\tau\,e^{i\Om_n\tau}e^{\tau(E_m-E_n)}
 =(p_n-p_m)/[\,p_m(i\Om_n+E_m-E_n)\,]$
[using $e^{i\Om_n\beta}=1$ and $e^{\beta(E_m-E_n)}=p_n/p_m$], together with the
transform $\beta\,\delta_{\Om_n,0}$ of the elastic constant, yields the
finite-temperature Lehmann function [$p_m=e^{-\beta(E_m-E_0)}$, $\mathcal Z=\sum_m p_m$]
\begin{equation}
\chi^{\mathrm{emb}}_{AB}(i\Om_n)
 = \frac1{\mathcal Z}\sum_{m\ne n}
 \frac{p_n-p_m}{\,i\Om_n-(E_n-E_m)\,}\,
 \langle m|A|n\rangle\langle n|B|m\rangle
 + \beta\,\delta_{\Om_n,0}\Big[\frac1{\mathcal Z}\sum_m p_m
 \langle m|A|m\rangle\langle m|B|m\rangle
 - \langle A\rangle\langle B\rangle\Big],
\label{eq:thermallehmann}
\end{equation}
with the degenerate static limit $(p_n-p_m)/[-(E_n-E_m)]\to +\beta p_m$ at $\Om_n=0$. The second term of
Eq.~\eqref{eq:thermallehmann} is the diagonal ($m{=}n$) elastic contribution,
present only at $\Om_n=0$.
All three embedding objects follow from the
\emph{same} function, Eq.~\eqref{eq:thermallehmann}, letting $A,B$ run over the channel
operator $\hat{\mathcal O}$ and the embedding fluctuation operators
$\partial\Hemb/\partial x_\mu$ [$\mu\in(D',D'',\lambda^c)$]: the bare channel
susceptibility is $\chi^{\mathrm{emb}}_{\hat{\mathcal O}\hat{\mathcal O}}$, the source
vectors are $\chi^{\mathrm{emb}}_{\hat{\mathcal O}x_\mu}$, and the embedding fluctuation
block is the $s=-1$ bubble of Eq.~\eqref{eq:secondvar},
\begin{equation}
\Memb_{\mu\nu}(i\Om_n) = -\,\chi^{\mathrm{emb}}_{x_\mu x_\nu}(i\Om_n) ,
\label{eq:Membchi}
\end{equation}
its diamagnetic contact vanishing because $\Hemb$ is linear in $(\mathcal D,\lambda^c)$.
Equation~(11) then reads
$\chi=\chi^{\mathrm{emb}}_{\hat{\mathcal O}\hat{\mathcal O}}
 +\sum_{\mu\nu}\chi^{\mathrm{emb}}_{\hat{\mathcal O}x_\mu}\,
 [\mathcal M^{-1}]_{\mu\nu}\,\chi^{\mathrm{emb}}_{x_\nu\hat{\mathcal O}}$
with $\mathcal M=\Mqp+\Mmix+\Memb$.

\section{Gauge transformation, gauge covariance, and gauge invariance}
\label{sec:gauge}

\emph{Gauge transformation.} The ghost extension introduces a gauge redundancy: the
physical content of the theory is unchanged under independent unitary rotations $u$
of the auxiliary quasiparticle ($f$) orbitals and $v$ of the bath ($b$) orbitals,
which act on the saddle-point fields as
\begin{equation}
R\to u^\dg R,\qquad \lambda\to u^\dg\lambda u,\qquad
\mathcal D\to v^{T}\mathcal D,\qquad \lambda^c\to v^\dg\lambda^c v .
\label{eq:gaugetrafo}
\end{equation}
Each Lagrangian term is invariant under Eq.~\eqref{eq:gaugetrafo}:
$\underline H^{\mathrm{qp}}\to u^\dg\,\underline H^{\mathrm{qp}}\,u$ leaves the
trace-log of Eq.~(4) unchanged, the transformation of $(\mathcal D,\lambda^c)$
amounts to a unitary rotation of the bath operators in Eq.~(8), and
$H^{\mathrm{emb}}_{0}$ of Eq.~\eqref{eq:buildH} is conjugated by
$\mathrm{diag}(u,v)$. Physical observables are invariant: for example,
$R^\dg R\to R^\dg u\,u^\dg R=R^\dg R$ and the physical propagator
of Eq.~\eqref{eq:Gphys} is unchanged. A given saddle point
$\bm x^{(0)}$, with fields $(R,\lambda,\mathcal D,\lambda^c)$, is therefore \emph{not} an isolated solution:
sweeping over all gauge rotations $g\equiv(u,v)$ of
Eq.~\eqref{eq:gaugetrafo} generates a continuous set of physically
equivalent saddle points,
\begin{equation}
g\!\cdot\!\bm x^{(0)} \equiv (u^\dg R,\; u^\dg\lambda u,\; v^{T}\mathcal D,\; v^\dg\lambda^c v),
\label{eq:gaugeorbit}
\end{equation}
which we call the \emph{gauge orbit} of $\bm x^{(0)}$ (the set of points reached from $\bm x^{(0)}$ by
the gauge rotations). As shown below, this redundancy makes the static fluctuation matrix
singular and dictates how the inverse in Eq.~(11) must be taken.

\emph{Construction of the transformation matrix $W$.} A gauge rotation of the
saddle induces a linear transformation of the fluctuation coordinates,
$\delta\bm x \to W\,\delta\bm x$, constructed group by group from the basis
expansion of Eq.~\eqref{eq:basis}. Writing the complex $R$-type coefficient vector
as $c^R_s=r'_s+i\,r''_s$ (and similarly $c^D_s$), the fluctuations transform as
$\delta R\to u^\dg\,\delta R$, $\delta\lambda\to u^\dg\,\delta\lambda\,u$,
$\delta\mathcal D\to v^{T}\delta\mathcal D$,
$\delta\lambda^c\to v^\dg\,\delta\lambda^c\,v$, so projecting back onto the
orthonormal bases $\{h_s^r,h_s\}$ gives the blocks
\begin{equation}
\begin{aligned}
\big[W^{R}\big]_{ss'} &= \Tr\!\big(h_s^{r\dg}\, u^\dg h_{s'}^r\big),&
\big[W^{\lambda}\big]_{ss'} &= \Tr\!\big(h_s^\dg\, u^\dg h_{s'} u\big),\\
\big[W^{D}\big]_{ss'} &= \Tr\!\big(h_s^{r\dg}\, v^{T} h_{s'}^r\big),&
\big[W^{\lambda^c}\big]_{ss'} &= \Tr\!\big(h_s^\dg\, v^\dg h_{s'} v\big),
\end{aligned}
\label{eq:Wblocks}
\end{equation}
with $c^R\to W^R c^R$, $l\to W^\lambda l$, etc. In the real fluctuation basis
$\bm x=(R',R'',\lambda,D',D'',\lambda^c)$, the complex block $W^{R}$ acts on the
$(R',R'')$ pair through the real matrix
\begin{equation}
\widetilde W^{R}=\begin{pmatrix} \mathrm{Re}\,W^R & -\mathrm{Im}\,W^R\\
 \mathrm{Im}\,W^R & \mathrm{Re}\,W^R\end{pmatrix},
\label{eq:Wreal}
\end{equation}
and likewise $W^{D}$ acts on the $(D',D'')$ pair through $\widetilde W^{D}$, while $W^\lambda$
and $W^{\lambda^c}$ are real orthogonal on the
Hermitian $\lambda$-type coefficients; the full $W$ is the block-diagonal
combination $W=\operatorname{diag}(\widetilde W^{R},W^{\lambda},\widetilde W^{D},W^{\lambda^c})$ and is unitary whenever the gauge rotation maps the fluctuation bases
$\{h_s^r,h_s\}$ onto themselves. All covariance statements below are restricted to rotations whose
action closes within the chosen charge or longitudinal-spin basis. Generic rotations that mix
transverse or imaginary-antisymmetric directions require the corresponding complete Hermitian
basis. In the $\zeta$-enlarged space of Sec.~\ref{sec:cqpa} the
constraint field is gauge neutral, $\zeta\to\zeta$ [both $V=R\mathcal O R^\dg$ and
$\hat{\mathcal O}$ transform covariantly], so $W$ is extended by a
trivial $1$ on the $\zeta$ coordinate.

\emph{Gauge covariance.} Denote the source vectors collectively by
$B_\mu$: $\chi^{\mathrm{emb}}_{x_\mu\mathcal O}$ for the embedding route or
$\chi^{\mathrm{GF}}_{x_\mu\mathcal O}$ (bubble \emph{plus} diamagnetic contact) for
the quasiparticle route. Under a gauge rotation of the saddle, $\bm x^{(0)}\to\bm x^{(0)\prime}$, the
fluctuation objects transform \emph{homogeneously},
\begin{equation}
\mathcal M(\bm x^{(0)\prime},\bm q,i\Om_n) = W\,\mathcal M(\bm x^{(0)},\bm q,i\Om_n)\,W^\dg,\qquad
B(\bm x^{(0)\prime},\bm q,i\Om_n) = W\,B(\bm x^{(0)},\bm q,i\Om_n),\qquad
\chi^{(0)}(\bm x^{(0)\prime}) = \chi^{(0)}(\bm x^{(0)}),
\label{eq:cov}
\end{equation}
which is the statement of gauge \emph{covariance}: the bilinear response formula,
Eq.~(11), is then gauge invariant in exact arithmetic because the $W$'s cancel
between $B^\dg$, $\mathcal M^{-1}$, and $B$. We verified Eq.~\eqref{eq:cov}
numerically for basis-preserving independent ($u\neq v$) ghost rotations: the sources and
$\chi^{(0)}$ transform to $\sim10^{-15}$ and $\mathcal M$ to
$10^{-8}$ or better.

\emph{Gauge invariance and null-space projection.} Gauge covariance makes
$\mathcal M(\bm q,0)$ singular. An infinitesimal gauge rotation
$g_\theta=(e^{i\theta\mathcal T_u},e^{i\theta\mathcal T_v})\simeq(\openone+i\theta\mathcal T_u,
\openone+i\theta\mathcal T_v)$ ($\mathcal T_{u},\mathcal T_{v}$ Hermitian)
maps the saddle $\bm x^{(0)}$ to the physically equivalent saddle
$\bm x^{\star}(\theta)\equiv g_\theta\!\cdot\!\bm x^{(0)}$,
along the orbit
of Eq.~\eqref{eq:gaugeorbit}: in the fluctuation basis of Eq.~\eqref{eq:basis}, its components
are the expansion coefficients of the rotated fields
$(u^\dg R,\;u^\dg\lambda u,\;v^{T}\mathcal D,\;v^\dg\lambda^c v)$ on $\{h_s^r,h_s\}$,
collected in the layout $\bm x=(R',R'',\lambda,D',D'',\lambda^c)$, with
$\bm x^{\star}(0)=\bm x^{(0)}$ the homogeneous saddle of Eq.~\eqref{eq:fourier}. Because $\mathcal L$ is gauge
invariant, every point of the orbit is itself a saddle, so the stationarity condition
holds identically in $\theta$,
\begin{equation}
\frac{\partial\mathcal L}{\partial x_\mu}\bigg|_{\bm x^{\star}(\theta)}=0
\qquad\text{for all }\theta .
\label{eq:statorbit}
\end{equation}
Differentiating this identity with respect to $\theta$, the chain rule gives
\begin{align}
0=\frac{d}{d\theta}\frac{\partial\mathcal L}{\partial x_\mu}\bigg|_{\bm x^{\star}(\theta)}
&=\sum_\nu\frac{\partial^2\mathcal L}{\partial x_\mu\partial x_\nu}\bigg|_{\bm x^{\star}(\theta)}\,
 \frac{d x^{\star}_\nu(\theta)}{d\theta}
\nonumber\\[2pt]
&\xrightarrow{\ \theta=0\ }\;
\sum_\nu\frac{\partial^2\mathcal L}{\partial x_\mu\partial x_\nu}\bigg|_{\bm x^{(0)}}\,
 [\delta_{\mathcal T}\bm x]_\nu
\;\Longleftrightarrow\;
\mathcal M(\bm q{=}0,\Om_n{=}0)\,\delta\bm x_{\mathcal T}=0 ,
\label{eq:zeromode}
\end{align}
where
$[\delta_{\mathcal T}\bm x]_\nu\equiv d x^{\star}_\nu(\theta)/d\theta\,\big|_{\theta=0}$ is the
orbit tangent, and the Hessian $\partial^2\mathcal L/\partial x_\mu\partial x_\nu$ is the
uniform static fluctuation matrix. The orbit tangent is thus an exact null vector:
moving along the gauge direction costs no action to second order.
The gauge symmetry is \emph{local}: the auxiliary frames may be rotated
independently at every site. A site-dependent $u_i$ is a change of basis of the
quasiparticle fermions, $\underline H^{\mathrm{qp}}\to U^\dg\underline
H^{\mathrm{qp}}U$ with $U=\bigoplus_i u_i$, which leaves the trace-log of Eq.~(4)
invariant; a site-dependent $v_i$ is a canonical rotation of the bath operators in
Eq.~(8); and $H^{\mathrm{emb}}_{0,i}$ of Eq.~\eqref{eq:buildH} is conjugated by
$\mathrm{diag}(u_i,v_i)$. The identity of Eq.~\eqref{eq:statorbit} therefore holds for
every Fourier component of the rotation angle, and Eq.~\eqref{eq:zeromode} gives
exact zero modes of $\mathcal M(\bm q,\Om_n{=}0)$ at \emph{every} $\bm q$.

Of the eighteen generators of $U(3)_f\times U(3)_b$ [nine $u$-type, acting on
$(R,\lambda)$, and nine $v$-type, acting on $(\mathcal D,\lambda^c)$; their
field-by-field variations are given in Eq.~\eqref{eq:orbittangents} of
Sec.~\ref{sec:cqpa}], the orbit tangents span a twelve-dimensional subspace of the
fluctuation space of Eq.~\eqref{eq:basis}. Take the $u$-type block; a Hermitian
$3\times3$ generator $t$ is either real symmetric (six of the nine) or imaginary
antisymmetric (the remaining three), and the channel generator
$\mathcal T=t\otimes\sigma^{0}$ (charge) or $t\otimes\sigma^{z}$ (spin) inherits
its symmetry type. For real symmetric $\mathcal T$, acting on the real saddle
fields, the variation $\delta\lambda=-i[\mathcal T,\lambda]$ is imaginary
antisymmetric and therefore orthogonal to every basis element
[$\Tr(h_s^\dg\delta\lambda)=0$, since $\Tr(SA)=0$ for symmetric $S$ and
antisymmetric $A$], while $\delta R=-i\,\mathcal T R$ is purely imaginary and
occupies only the three $R''$ coordinates; six such tangents confined to three
coordinates span only three directions. The three imaginary antisymmetric
generators contribute a real $\delta R$ and a real-symmetric $\delta\lambda$,
adding three more. The nine $u$-type generators therefore span a six-dimensional
tangent space, and the $v$-type block counts identically: twelve exact zero modes
per channel at every $\bm q$. The three remaining generator combinations per block
vary the saddle only along directions outside the charge and longitudinal-spin
fluctuation space, namely directions that belong to the complex basis of the
transverse spin channels (Sec.~\ref{sec:lagrangian}), so they contribute no
mode to $\mathcal M$ and no response in the channels computed here; they appear as
the six vanishing singular values in Sec.~\ref{sec:cqpa}. Since the zero-mode
directions carry no physical static response, the inverse $\mathcal M^{-1}$ in Eq.~(11) must be
restricted to the complementary subspace. We therefore project onto the orthonormal complement
$w_{i\mu}$ of the zero-mode (null) space of the static Hessian $\mathcal M(\bm q,\Om_n{=}0)$,
\begin{align}
\bar{\mathcal M}_{ij}(\bm q,i\Om_n)=\sum_{\mu\nu} w_{i\mu}\,\mathcal M_{\mu\nu}(\bm q, i\Om_n)\,w_{j\nu},\quad
\bar\chi_{i\mathcal O}(i\Om_n)=\sum_\mu w_{i\mu}\,\chi^{\mathrm{emb}}_{x_\mu\mathcal O}(i\Om_n),
\label{eq:projdef} \\
\chi_{\mathcal O\mathcal O}(\bm q, i\Om_n) = \chi^{\mathrm{emb}}_{\mathcal O\mathcal O}(i\Om_n)
 + \sum_{ij}\bar\chi_{i\mathcal O}(i\Om_n)\,\big[\bar{\mathcal M}^{-1}\big]_{ij}(\bm q, i\Om_n)\,
   \bar\chi_{j\mathcal O}(i\Om_n).
\label{eq:chiproj}
\end{align}
The projector is built \emph{once} from the static Hessian, and the same $w$ is applied at
every frequency. At $\Om_n=0$ this removes exactly the null space. At finite frequency the static
zero-mode directions are in general no longer null vectors of $\mathcal M(\bm q,i\Om_n)$, so
applying the same projector there is a gauge-fixing prescription. The invariance mechanism is transparent from Eq.~\eqref{eq:cov}: under a basis-preserving
static gauge rotation, the static Hessian
$\mathcal M(\bm x^{(0)\prime})=W\mathcal M(\bm x^{(0)})W^\dg$ has the \emph{same} eigenvalues as
$\mathcal M(\bm x^{(0)})$, so the kept eigenvector subspace transforms as
$w(\bm x^{(0)\prime})=W w(\bm x^{(0)})$ (up to an irrelevant unitary rotation within that subspace).
Because Eq.~\eqref{eq:cov} holds at every frequency, the $W$'s then cancel in the reduced objects
$\bar{\mathcal M}$ and $\bar\chi$ of Eq.~\eqref{eq:projdef}, leaving them, and hence
$\chi_{\mathcal O\mathcal O}$ of Eq.~\eqref{eq:chiproj}, gauge invariant for arbitrary
$(\bm q,i\Om_n)$. This invariance is verified numerically below.

\emph{Numerical construction of the projector.} The projector is built from the
singular-value (equivalently, at $\Om_n=0$, eigenvalue) decomposition of the static
fluctuation matrix. At $\Om_n=0$ the Hessian $\mathcal M(\bm q,0)$ is Hermitian; we
diagonalize it and discard the gauge subspace with a \emph{threshold} $\eta_0\sim10^{-6}$,
keeping the eigenvectors with $|\text{eigenvalue}|>\eta_0$.
The kept eigenvectors form the columns $w_{i\mu}$ of
Eq.~\eqref{eq:projdef}, an orthonormal basis of the physical subspace. This
\emph{static} projector is used throughout: the same $w$ is applied at every
frequency, $\bar{\mathcal M}(i\Om_n)=w^\dg\,\mathcal M(i\Om_n)\,w$, and the
resulting embedding-source susceptibility is gauge invariant to $10^{-10}$ or
better at every Matsubara frequency in both channels.

\section{Conserving quasiparticle approximation: \texorpdfstring{$\zeta$}{zeta} blocks and zero-mode projection}
\label{sec:cqpa}

\emph{The $\zeta$ constraint.} Attaching the source
to the quasiparticle sector alone is non-conserving: the source couples to the
density built from the quasiparticle Green's function, while the interaction
physics enters through the embedding density, and nothing constrains the two
responses to agree. The dynamical field $\zeta_i(\tau)$ enforces the missing
consistency by equating the two local densities,
\begin{equation}
\Tr\!\big[\mathcal O\,G_{ii}(\tau,\tau^{+})\big]
 = \big\langle\hat{\mathcal O}_{i}(\tau)\big\rangle_{\rm emb} ,
\label{eq:zetaexplicit}
\end{equation}
with $G_{ij}(\tau,\tau')=R^\dg_i(\tau)\,\underline G_{ij}(\tau,\tau')\,R_j(\tau')$
the physical Green's function built from the quasiparticle one $\underline G$ of
the fluctuating configuration, and $\langle\cdot\rangle_{\rm emb}$ the
imaginary-time-ordered embedding average. Reduced to a single Fourier block as in
Sec.~\ref{sec:derivation}, Eq.~\eqref{eq:zetaexplicit} is imposed independently on
each Fourier component by $\zeta_{\bm q,\Om_n}$.

The corresponding background densities at the saddle do not match exactly: the
quasiparticle side carries total weight $R^\dg R<1$, whereas the embedding average
carries the full physical weight, so their difference is controlled by the
spectral-weight deficit $1-R^\dg R$. Rather than reoptimizing the static saddle,
we retain the original gRISB solution with $\zeta^{(0)}=0$ and impose
Eq.~\eqref{eq:zetaexplicit} in the Gaussian fluctuation theory. Thus, the density
equality is the constraint being enforced, while its realization about the finite-bath
background is approximate. We quantify this approximation by the relative difference between the
background quasiparticle and embedding density matrices,
which tracks the deficit: at $n=0.75$ it is $2.5\%$ for gRISB
($N_a=6$, $U=10$, $R^\dg R=0.944$) against $30\%$ for RISB ($N_a=2$, $U=10$,
$R^\dg R=0.695$), and falls to $0.5\%$ once the deficit is small ($N_a=2$, $U=1$,
$R^\dg R=0.995$). The ghost orbitals therefore do more than sharpen the spectrum:
by restoring the missing spectral weight they make $\zeta=0$ a controlled mean-field
starting point, which the RISB limit does not provide.

\emph{Quasiparticle source and the $\zeta$ multiplier.} The conserving
quasiparticle approximation follows from the construction of
Sec.~\ref{sec:derivation} with the source moved to the quasiparticle sector. There
the physical operator is the $R$-dressed bilinear with vertex
$R_i(\tau)\,\mathcal O\,R^\dg_i(\tau)$, so, since
$\Hqp_{\bm k}=R\,\eps_{\bm k}R^\dg+\lambda$, it enters as a $\lambda$-type shift.
In place of Eq.~\eqref{eq:source}, the source $\eta$ and the multiplier $\zeta$
enforcing the density constraint of Eq.~\eqref{eq:zetaexplicit} therefore act with
opposite signs as
\begin{equation}
\lambda_i(\tau)\;\to\;\lambda_i(\tau)
 +\big[\zeta_i(\tau)-\eta_i(\tau)\big]\,R_i(\tau)\,\mathcal O\,R^\dg_i(\tau) ,
\qquad
\Hemb_i(\tau)\;\to\;\Hemb_i(\tau)-\zeta_i(\tau)\,\hat{\mathcal O}_i .
\label{eq:cqpasource}
\end{equation}
Equations~\eqref{eq:Ldyn}--\eqref{eq:Lparts} then read
\begin{equation}
\begin{aligned}
\mathcal L_{\mathrm{qp}}[\bm x,\zeta;\eta] &=
 -\frac{1}{\beta N}\ln\Tr\,\mathcal T_\tau
  \exp\!\Big[-\!\int_0^{\beta}\! d\tau\,\hat H^{\mathrm{qp}}
  \big(\{R_i(\tau),\,\lambda_i(\tau)
  +[\zeta_i(\tau)-\eta_i(\tau)]\,R_i(\tau)\,\mathcal O\,R^\dg_i(\tau)\}\big)\Big],\\
\mathcal L_{\mathrm{emb}}[\bm x;\zeta] &=
 -\frac{1}{\beta N}\sum_i\ln\Tr\,\mathcal T_\tau
  \exp\!\Big[-\!\int_0^{\beta}\! d\tau\,\big(\Hemb_i(\tau)
  -\zeta_i(\tau)\,\hat{\mathcal O}_i\big)\Big],
\end{aligned}
\label{eq:cqpaL}
\end{equation}
with $\mathcal L_{\mathrm{mix}}$ unchanged. The vertex carries the fluctuating
$R_i(\tau)$, which is the origin of the diamagnetic term in
Eq.~\eqref{eq:chixOgf} below. At finite bath size, the couplings in
Eq.~\eqref{eq:cqpaL} generate a uniform term linear in $\zeta$ at the saddle point, equal to the small
background density mismatch discussed above. Keeping $\zeta^{(0)}=0$ amounts to
neglecting this residual first derivative while retaining all quadratic $\zeta$
blocks and source vertices. The residual does not enter the Gaussian Hessian and
vanishes in the complete-bath limit. The quadratic response about this approximate
constrained saddle is therefore evaluated in the enlarged space
$\bm x\to(\bm x,\zeta)$, giving
\begin{equation}
\chi_{\mathcal O\mathcal O}(\bm q,i\Om_n)
 = \chi^{(0)}_{\mathcal O\mathcal O}(\bm q,i\Om_n)
 + \sum_{\mu\nu}\gamma_\mu(-\bm q,-i\Om_n)\,
   \big[\tilde{\mathcal M}_\zeta(\bm q,i\Om_n)\big]^{-1}_{\mu\nu}\,
   \gamma_\nu(\bm q,i\Om_n) ,
\label{eq:cqpachain}
\end{equation}
with $\mu,\nu$ running over $(\bm x,\zeta)$; these are Eqs.~(15)--(17) of the main text.
Since $\mathcal M_\zeta$ inherits the static gauge zero modes of Sec.~\ref{sec:gauge},
$\tilde{\mathcal M}_\zeta^{-1}$ denotes the pseudo-inverse on the physical subspace,
constructed below. The two legs are inequivalent at finite $\Om_n$, the
left one carrying the reversed transfer as written.
The bare term is the second derivative of $\mathcal L_{\mathrm{qp}}$ in the source,
and the legs are its mixed derivatives in the source and each coordinate:
$\gamma_\mu\equiv\chi^{\mathrm{GF}}_{x_\mu\mathcal O}$ on the quasiparticle directions,
$\gamma_\mu=0$ on the embedding directions since $\eta$ does not enter
$\mathcal L_{\mathrm{emb}}$, and
$\gamma_\zeta\equiv\chi^{(0)}_{\mathcal O\mathcal O}$ on the constraint direction,
because $\eta$ and $\zeta$ enter $\mathcal L_{\mathrm{qp}}$ only through the
combination $(\zeta-\eta)$. %

\emph{Quasiparticle-source ingredients.} The bare susceptibility and the source
vectors appearing in Eq.~\eqref{eq:cqpachain} are lattice quasiparticle bubbles,
\begin{align}
\chi^{(0)}_{\mathcal O\mathcal O}(\bm q,i\Om_n)
 &= -\frac{1}{\beta N}\sum_{\bm k,\omega_n}
 \Tr\!\Big[R^\dg\,\underline G_{\bm k+\bm q}(i\omega_n{+}i\Om_n)\,R\,\mathcal O\,
 R^\dg\,\underline G_{\bm k}(i\omega_n)\,R\,\mathcal O\Big],
\label{eq:chi0gf}\\
\chi^{\mathrm{GF}}_{x_\mu\mathcal O}(\bm q,i\Om_n)
 &= -\frac{1}{\beta N}\sum_{\bm k,\omega_n}
 \Tr\!\Big[\underline G_{\bm k}(i\omega_n)\,\Lambda^{\bm k,-\bm q}_\mu\,
 \underline G_{\bm k+\bm q}(i\omega_n{+}i\Om_n)\,R\,\mathcal O R^\dg\Big]
 \;-\;\frac1N\sum_{\bm k}
 \tr\!\Big[\nF(\Hqp_{\bm k})\,
 \frac{\partial(R\,\mathcal O R^\dg)}{\partial x_\mu}\Big],
\label{eq:chixOgf}
\end{align}
with the same vertices, Eq.~\eqref{eq:vertices}, as $\Mqp$. The second (diamagnetic) term
of Eq.~\eqref{eq:chixOgf} is the one anticipated above, from the $R$ dependence of the
vertex in Eq.~\eqref{eq:cqpasource}; it is nonzero only for the $R',R''$ directions and
\emph{must} be kept: it is required for gauge covariance of the source.

\emph{$\zeta$ blocks of the enlarged fluctuation matrix.} The fluctuation space is
enlarged by one coordinate per channel, $\bm x\to(\bm x,\zeta)$,
$N_M\to N_M+1$, and $\mathcal M\to\mathcal M_\zeta$. Because $\zeta$ enters
$\underline H^{\mathrm{qp}}$ exactly as a $\lambda$-type direction with vertex
$V=R\,\mathcal O R^\dg$, its quasiparticle blocks follow from the same functional
derivatives as the $\lambda$ rows of $\Mqp$ with the substitution $h_s\to V$; no
separate ansatz is needed. Explicitly,
\begin{equation}
[\mathcal M_\zeta]_{x_\mu \zeta}(\bm q,i\Om_n)
 = -\,\chi^{\mathrm{GF}}_{x_\mu \mathcal O}(\bm q,i\Om_n)
 \;+\;\chi^{\mathrm{emb}}_{x_\mu\,\mathcal O}(i\Om_n),
\qquad
[\mathcal M_\zeta]_{\zeta\zeta}(\bm q,i\Om_n)
 = -\,\chi^{(0)}_{\mathcal O\mathcal O}(\bm q,i\Om_n)
 \;-\;\chi^{\mathrm{emb}}_{\mathcal O\mathcal O}(i\Om_n),
\label{eq:zetablocks}
\end{equation}
i.e.\ each $\zeta$ row/column is the sum of a quasiparticle strip
[Eq.~\eqref{eq:chixOgf} evaluated with the vertex $V$, from the $+\zeta V$ coupling]
and an embedding Lehmann strip
[Eq.~\eqref{eq:thermallehmann} with the $-\zeta\hat{\mathcal O}$
coupling]. The opposite-sign couplings give the two strips of the cross element opposite signs, while both contributions to the $\zeta\zeta$ element are negative because each coupling enters twice.
Here $\chi^{\mathrm{emb}}_{x_\mu\mathcal O}$ denotes the Lehmann response
between $\partial\Hemb/\partial x_\mu$ and $\hat{\mathcal O}$,
nonzero only on the embedding directions, while
$\chi^{\mathrm{GF}}_{x_\mu\mathcal O}$ is nonzero only on the quasiparticle
directions.
This completes the ingredients of Eq.~\eqref{eq:cqpachain}.

\emph{Zero-mode projection for the quasiparticle source.} We can utilize the projection procedure of Sec.~\ref{sec:gauge} to remove the gauge zero modes. Here, we present an alternative way to build the projector
analytically from the gauge group generators, evaluated directly from the saddle
fields. For channel generators $\mathcal T=t\otimes\sigma^0$ (charge) or
$\mathcal T=t\otimes\sigma^z$ (spin), the two independent rotations
$u=e^{i\theta\mathcal T}$ and $v=e^{i\theta\mathcal T}$ of
Eq.~\eqref{eq:gaugetrafo} change the saddle fields by
\begin{equation}
\begin{gathered}
u\text{-type}:\quad \delta R = -i\,\mathcal T R,\qquad
\delta \lambda = -i\,[\mathcal T,\lambda],\qquad
\delta \mathcal D = 0,\qquad \delta \lambda^c = 0,\\[3pt]
v\text{-type}:\quad \delta R = 0,\qquad \delta \lambda = 0,\qquad
\delta \mathcal D = +i\,\mathcal T^{T} \mathcal D,\qquad
\delta \lambda^c = -i\,[\mathcal T,\lambda^c],\qquad \delta\zeta = 0 ,
\end{gathered}
\label{eq:orbittangents}
\end{equation}
with $t$ running over the nine Hermitian generators of $U(3)$ in each of the two
sectors (per spin and charge channel) for the one-band $N_a=6$ model; these are the zero-mode
variations $\delta\bm x_{\mathcal T}$ of Eq.~\eqref{eq:zeromode}, written out field
by field.

\emph{Construction of the projector.} The gauge-mode projector is built once per
channel from the variations of Eq.~\eqref{eq:orbittangents}. For each generator $t$ we
evaluate the variations $(\delta R,\delta\lambda,\delta\mathcal D,\delta\lambda^c,\delta\zeta)$
at the homogeneous saddle and expand them in the fluctuation basis of
Eq.~\eqref{eq:basis}. The complex coefficients $\Tr(h_s^{r\dg}\,\delta R)$ and
$\Tr(h_s^{r\dg}\,\delta\mathcal D)$ give the $(R',R'')$ and $(D',D'')$ coordinates through
their real and imaginary parts, and the real coefficients $\Tr(h_s^\dg\,\delta\lambda)$ and
$\Tr(h_s^\dg\,\delta\lambda^c)$ give the $\lambda$ and $\lambda^c$ coordinates. Together with
$\delta\zeta=0$, they form one column vector in the
$(R',R'',\lambda,D',D'',\lambda^c,\zeta)$ layout of $\bm x$. The eighteen column vectors,
one per generator, form the matrix $K\in\mathbb{R}^{(N_M+1)\times 18}$, whose singular values show
a clean gap fixing the rank at $r=12$ (Sec.~\ref{sec:gauge}); the first $r$ left
singular vectors $U_{\mathrm{orb}}$ span the gauge-mode space, and
$P_{\mathrm{orb}}=\openone-U_{\mathrm{orb}}U_{\mathrm{orb}}^{T}$ projects onto its
orthogonal complement. Both the fluctuation matrix and the source vectors are
projected,
$\tilde{\mathcal M}_\zeta = P_{\mathrm{orb}}\mathcal M_\zeta P_{\mathrm{orb}}$ and
$\bar B = P_{\mathrm{orb}} B$, before the inversion in Eq.~(15).

\section{Numerical details and validation}
\label{sec:numerics}

\emph{Setup.} All calculations use the two-dimensional one-band Hubbard model,
Eq.~(1), on an $N_k=128^2$ momentum grid, with $N_a=6$ ($N_a=2$ for the RISB
comparisons). Saddle points are converged to $10^{-8}$ or better. The dynamical
susceptibilities of Fig.~3 are evaluated directly on the real-frequency axis by the analytic
continuation $i\Om_n\to\omega+i\eta$ [$\eta=0.05$, the spectral broadening,
distinct from the source $\eta_i(\tau)$ of Sec.~\ref{sec:derivation} and the
threshold $\eta_0$ of Sec.~\ref{sec:gauge}] applied in the Lindhard and
Lehmann kernels, on the path $\Gamma$--$X$--$M$--$\Gamma$ (61 points) with
$\omega\in[0,15]$; the Matsubara-axis results of Fig.~4 use
$\Om_n=2\pi nT$, $n=0,\dots,40$. For the real-frequency evaluations the static
projectors of Secs.~\ref{sec:gauge} and \ref{sec:cqpa} are built from the exact
$i\Om_n=0$, $\eta=0$ Hessian, and the same projector is applied on the
$\omega+i\eta$ grid.

\emph{Operator normalization.} The charge and spin channel operators are
$\mathcal O=\sigma^0/\sqrt2$ and $\sigma^z/\sqrt2$, i.e.\
$\chi_{ch}=\tfrac12\chi_{nn}$ and $\chi_{sp}=\tfrac12\chi_{\sigma^z\sigma^z}$ in
terms of the density and spin-density response.

\emph{Magnetic phase boundary (Fig.~1).} The boundary is located by the divergence of
the static spin susceptibility: at fixed $U$, the filling at which
$1/\chi_{sp}(\bm q^*,0)$ changes sign, where $\bm q^*$ is the leading instability
momentum along the $\Gamma$--$X$--$M$--$\Gamma$ path. The filling scans use
steps $\Delta n=0.025$, refined near the boundary (down to $\Delta n=10^{-3}$).

\emph{Validation of the Matsubara susceptibility.} Three routes were computed in
the same $\zeta$-enlarged theory at $U=10$, $n=0.75$, $\beta=50$,
$\bm q\in\{(0,0),(\pi,0),(\pi,\pi),(\pi,0.7\pi)\}$, $\Om_n=2\pi nT$,
$n=0,\dots,40$: (1) the embedding source without $\zeta$, (2) the embedding source
with $\zeta$, and (3) the CQPA. Adding $\zeta$ leaves the embedding-source
susceptibility unchanged to $4\times10^{-3}$ at the incommensurate spin peak
at $\bm q=(\pi,0.7\pi)$, where $\chi_{sp}\simeq5.1$, and to $4\times10^{-4}$ in the tails, confirming that the
enlargement is consistent. The CQPA agrees with the embedding source to
$|\chi^{\mathrm{CQPA}}-\chi|\le5.4\times10^{-3}$ (spin) and
$7\times10^{-4}$ (charge) at every momentum and Matsubara frequency, including the
near-critical incommensurate $\bm q=(\pi,0.7\pi)$; relative to the peak of each
channel this is $0.11\%$ for the spin response and $1.0\%$ for the charge
response, whose peak value is two orders of magnitude smaller. At $\bm q=(0,0)$
the CQPA gives $\chi(\Om_n\neq0)=0$ to machine precision
($\le5\times10^{-15}$ in both channels), the conservation law enforced by
$\zeta$. %

\emph{Finite-difference validation (Fig.~2, red diamonds).} The static
susceptibilities at the commensurate momenta were checked against field-derivative
calculations at the saddle-point level, independent of the entire fluctuation
machinery. A field term $\tfrac{f}{2}\sum_i \epsilon_i\,\hat{\mathcal O}_i$ (with
$\hat{\mathcal O}_i$ the local charge or spin operator in the normalization above,
and $\epsilon_i$ the pattern $+1$ at $\bm q=\Gamma$, striped for $X$, checkerboard
for $M$) is added to the Hamiltonian, the saddle point is re-converged at $\pm f$,
and $\chi=\delta\langle\mathcal O\rangle/f$ is formed by central differences, with
$f=5\times10^{-4}$ (spin) and $4\times10^{-3}$ (charge). The uniform ($\Gamma$)
response is computed in the single-site theory: grand canonically (fixed chemical
potential) for the charge channel and canonically (fixed density) for the spin
channel; the staggered $X$ and $M$ responses use two-site supercells with the
correspondingly folded dispersions on the full original Brillouin zone. Table
\ref{tab:fd} compares the two: the agreement is at the percent level or better,
limited by the different momentum grids of the folded supercell and the fluctuation
calculation.

\begin{table}[t]
\begin{ruledtabular}
\begin{tabular}{lcccccc}
 & \multicolumn{3}{c}{$\chi_{sp}$} & \multicolumn{3}{c}{$\chi_{ch}$} \\
 & $\Gamma$ & $X$ & $M$ & $\Gamma$ & $X$ & $M$ \\
\hline
fluctuation [Eq.~(11)]      & 0.6440 & 0.9286 & 1.0062 & 0.0665 & 0.0628 & 0.0673 \\
finite difference        & 0.6440 & 0.9386 & 1.0066 & 0.0665 & 0.0629 & 0.0673 \\
\end{tabular}
\end{ruledtabular}
\caption{Static spin and charge susceptibilities at the commensurate momenta from
the fluctuation formalism versus saddle-point finite-difference (field-derivative)
calculations, at $U=10$, $n=0.75$, $\beta=50$.\label{tab:fd}}
\end{table}

\begin{figure}[t]
\begin{centering}
\includegraphics[scale=0.46]{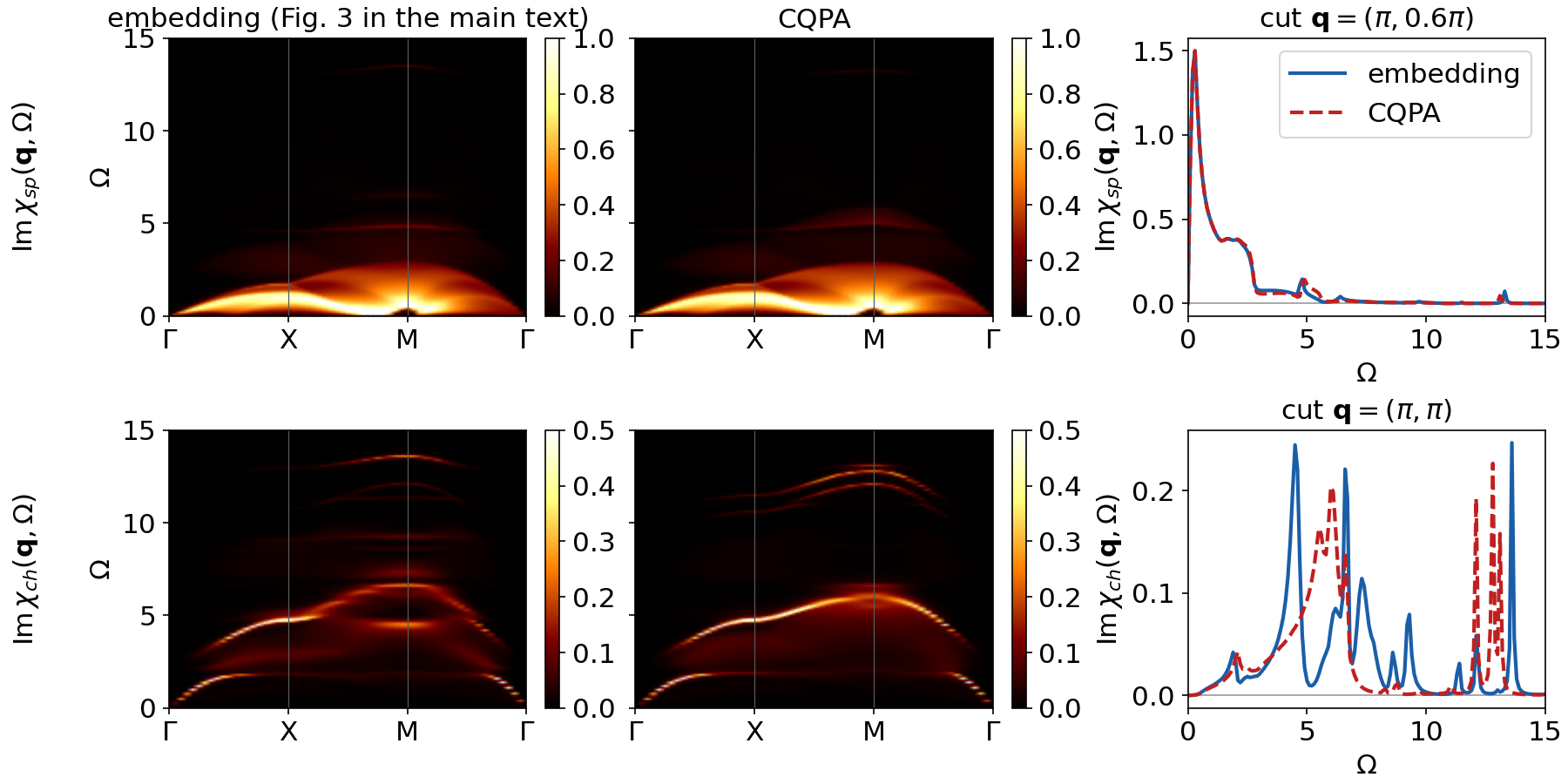}
\par\end{centering}
\caption{(Left column) The gRISB dynamical real-frequency spin susceptibility $\mathrm{Im}\,\chi_{sp}(\mathbf{q},\Omega)$ (top) and charge susceptibility (bottom) computed from Eq.~(11) of the main text. (Middle column) The gRISB dynamical real-frequency spin susceptibility $\mathrm{Im}\,\chi_{sp}(\mathbf{q},\Omega)$ (top) and charge susceptibility (bottom) computed from the conserving quasiparticle approximation (CQPA). (Right column) The spin and charge susceptibilities along selected momentum cuts. Here $U=10$, $n=0.75$, and $\beta=50$. \label{fig:cqpa_vs_emb}}
\end{figure}

\emph{Dynamical susceptibility: embedding formalism vs.\ conserving quasiparticle approximation.} We compare the real-frequency dynamical susceptibility obtained from the embedding formalism [Eq.~(11) of the main text] with that from the conserving quasiparticle approximation (CQPA) in Fig.~\ref{fig:cqpa_vs_emb}. The spectral line shapes obtained from the two approaches differ, most visibly in the charge channel, where the peak positions and weights also shift. The major excitations, however, are captured qualitatively by the CQPA: the paramagnon in the spin channel, and the sound mode and the Hubbard excitations in the charge channel. These results indicate that the additional quasiparticle constraint in the CQPA modifies the detailed dynamics while retaining the major physical excitations at a qualitative level.

\emph{Additional real-frequency dynamical charge susceptibility data.} We present additional data for the real-frequency dynamical charge susceptibility, $\mathrm{Im}\,\chi_{ch}(\mathbf{q},\Omega)$, at several electron fillings at $U=10$ and $\beta=50$, shown in Fig.~\ref{fig:chi_ch_dos}. Our RISB results (the left column of Fig.~\ref{fig:chi_ch_dos}) agree with the previous Kotliar--Ruckenstein slave-boson dynamical charge susceptibility~\cite{Riegler2020}, demonstrating the equivalence of the two response formalisms. The gRISB charge susceptibility shows additional excitations compared to the RISB susceptibility due to the incoherent scattering between the quasiparticles and Hubbard excitations. Moreover, at half filling, $n=1$, we observe a charge gap within the gRISB charge response, consistent with the Mott-insulating density of states shown in the right column of Fig.~\ref{fig:chi_ch_dos}.

\begin{figure}[t]
\begin{centering}
\includegraphics[scale=0.43]{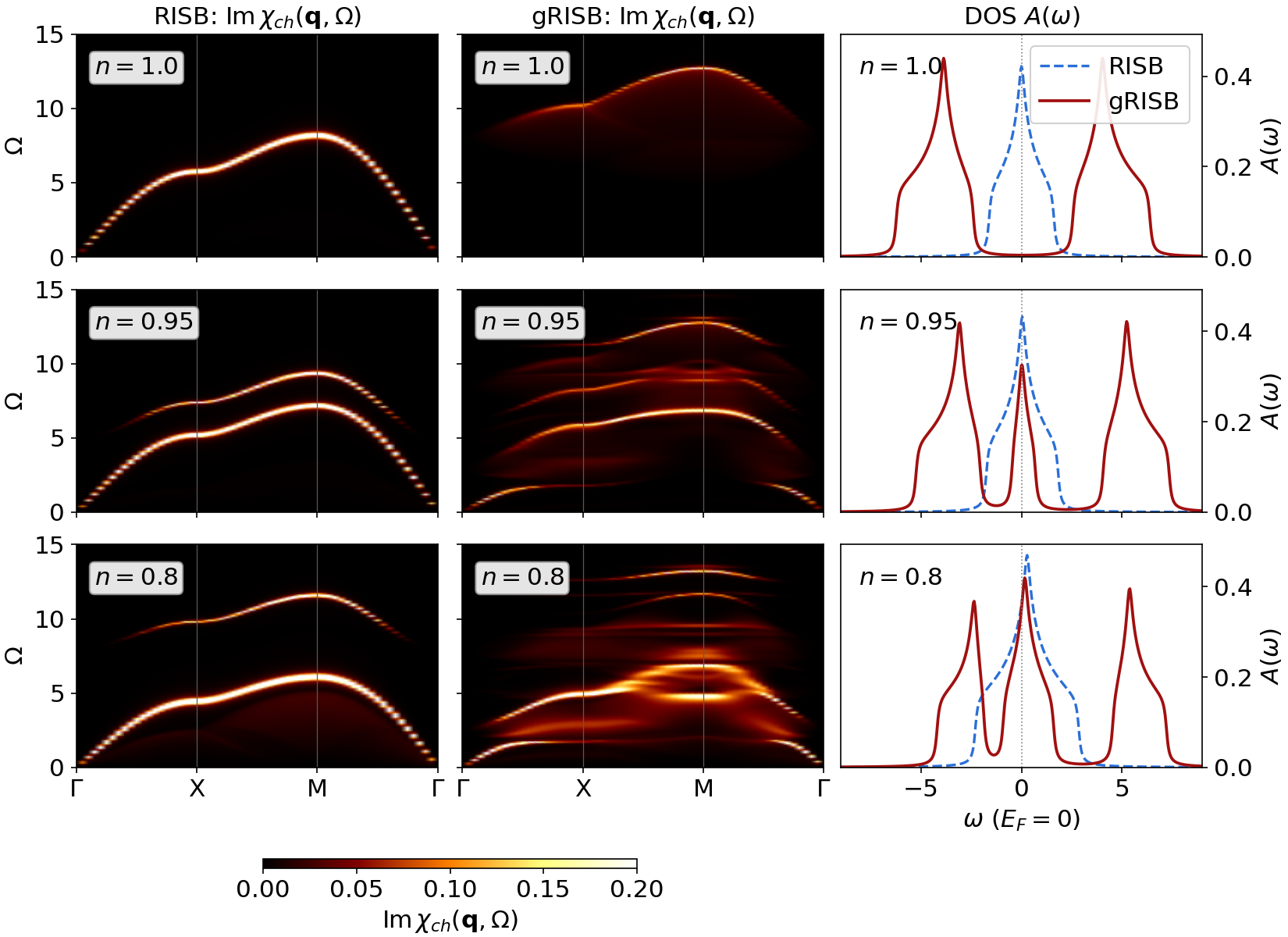}
\par\end{centering}
\caption{(Left column) The RISB charge susceptibility $\mathrm{Im}\,\chi_{ch}(\mathbf{q},\Omega)$ for electron filling $n=1,\ 0.95,\ 0.8$. (Middle column) The gRISB charge susceptibility $\mathrm{Im}\,\chi_{ch}(\mathbf{q},\Omega)$ for electron filling $n=1,\ 0.95,\ 0.8$. (Right column) The RISB and gRISB density of states (DOS). The Coulomb interaction is $U=10$, and the inverse temperature is $\beta=50$. \label{fig:chi_ch_dos}}
\end{figure}

\begin{figure}[t]
\begin{centering}
\includegraphics[width=0.98\linewidth]{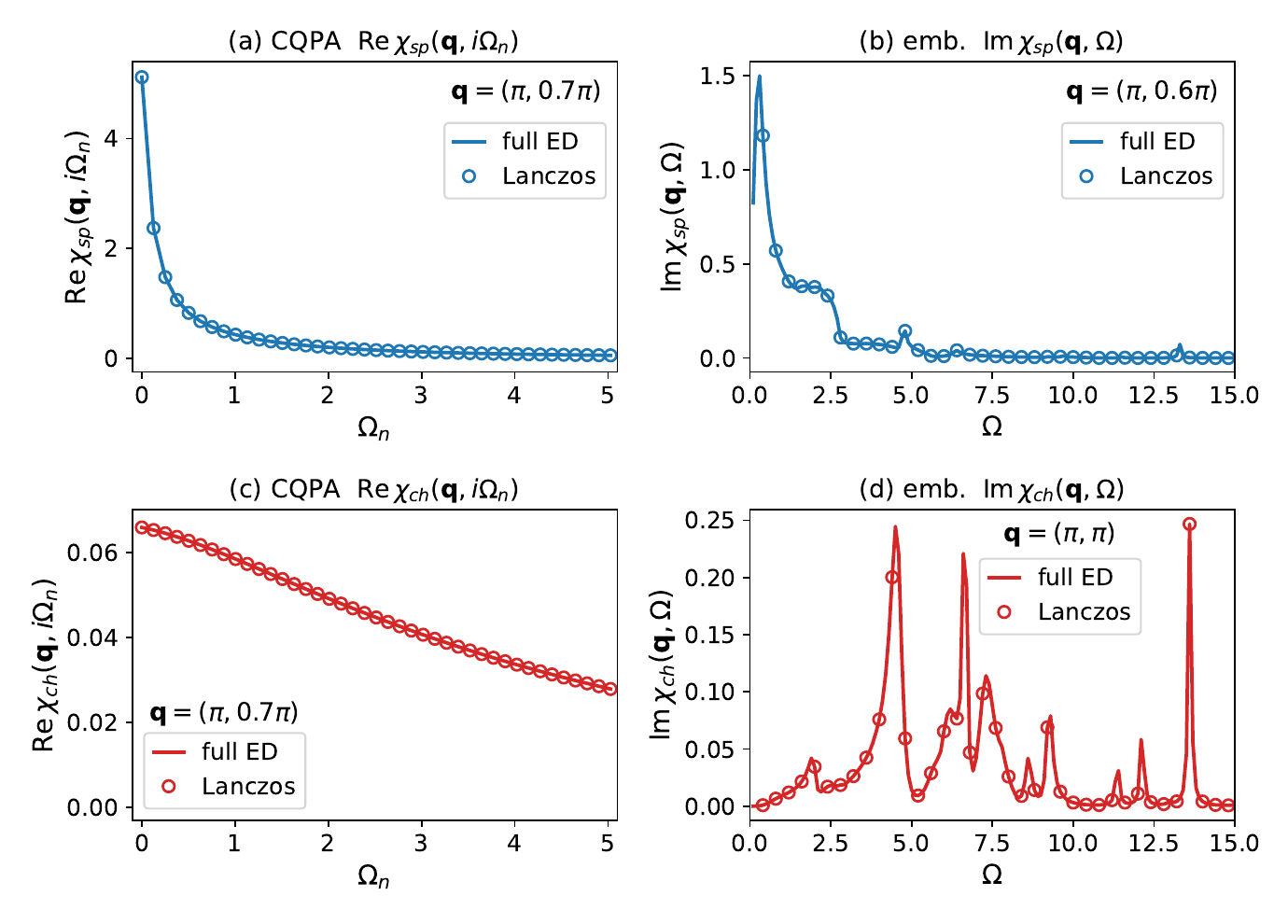}
\par\end{centering}
\caption{Continued-fraction Lanczos (open markers) versus full exact
diagonalization (solid lines) of the embedding Hamiltonian, gRISB $N_a=6$,
$U=10$, $n=0.75$, $\beta=50$. Left: CQPA susceptibilities on the Matsubara axis,
(a) spin and (c) charge [cf.\ Fig.~4]. Right: embedding-formalism susceptibilities
on the real-frequency axis, (b) spin and (d) charge [cf.\ Fig.~3]. The two solvers
agree to better than $10^{-6}$.\label{fig:lanczos}}
\end{figure}

\section{Continued-fraction Lanczos embedding solver}
\label{sec:lanczos}

The embedding correlators $\chi^{\mathrm{emb}}_{AB}$ of
Eq.~\eqref{eq:embcorr} are the only quantities in the response that require
diagonalizing $\Hemb$. At low temperature the thermal trace is dominated by the
ground state $|0\rangle$ of $\Hemb$ with ground-state energy $E_0$, and
$\chi^{\mathrm{emb}}_{AB}$ can be approximated by its zero-temperature (Lehmann) form
\begin{equation}
\chi^{\mathrm{emb}}_{AB}(z)=
 \langle 0|\,B\,\frac{1}{z+(\Hemb-E_0)}\,A\,|0\rangle
-\langle 0|\,A\,\frac{1}{z-(\Hemb-E_0)}\,B\,|0\rangle ,
\label{eq:cf_corr}
\end{equation}
with $z=i\Om_n$ (Matsubara) or $\omega+i\eta$ (retarded) and
$A,B\in\{\hat{\mathcal O},\,\partial\Hemb/\partial x_\mu\}$ as in
Eq.~\eqref{eq:embcorr}. Each matrix element follows from the Lanczos continued
fraction of a start vector $|\phi\rangle$,
\begin{equation}
g_\phi(z)\equiv\langle\phi|\,\frac{1}{z-(\Hemb-E_0)}\,|\phi\rangle=
 \cfrac{\langle\phi|\phi\rangle}
 {z-a_0-\cfrac{b_1^{2}}{z-a_1-\cfrac{b_2^{2}}{z-a_2-\cdots}}} ,
\label{eq:cf}
\end{equation}
with $\{a_n,b_n\}$ obtained by tridiagonalizing $\Hemb-E_0$ in the Krylov space
of $|\phi\rangle$; the general matrix element then follows by polarization,
\begin{equation}
\langle 0|A\,\frac{1}{z-(\Hemb-E_0)}\,B|0\rangle
=\frac14\!\!\sum_{s\in\{\pm1,\pm i\}}\!\!\bar s\,g_{\phi_s}(z),\quad
|\phi_s\rangle=A^\dg|0\rangle+s\,B|0\rangle .
\label{eq:cfAB}
\end{equation}
The full spectrum of
$\Hemb$ enters through the continued fraction, so only the ground state, obtained
by the same Lanczos recursion, and a few hundred Krylov vectors are required, in
place of a complete diagonalization. This makes the response accessible for the
larger embedding Hilbert spaces of multiorbital models, and dovetails with
compressed and quantum embedding solvers~\cite{Yao2021,PCA2026,Besserve2022,Chen2026}.

Figure~\ref{fig:lanczos} benchmarks the continued-fraction Lanczos against
full exact diagonalization for $N_a=6$: the CQPA susceptibilities on the Matsubara
axis (left) and the embedding-formalism susceptibilities on the real-frequency axis
(right) computed by the two solvers are indistinguishable, reproducing Figs.~4 and
3 of the main text, respectively.

\section{Computing the response functions in DMFT}
\label{sec:dmft}

The DMFT response functions are computed using \textit{Portobello}~\cite{adler_portobello_2024} and \textit{ComCTQMC}~\cite{melnick_accelerated_2021}, which employs worm measurements~\cite{gunacker_continuous-time_2015, gunacker_worm-improved_2016} of the one-, two-, and three-time local susceptibilities in order to compute the full vertex and its asymptotics~\cite{kaufmann_continuous-time_2017}. The worm measurements are taken after converging the DMFT equations. The asymptotic full vertex, $F_{a}$, and directly measured full vertex, $F_{m}$, are combined into the full vertex, $F$, as follows:
\begin{align}
F(i\nu,i\nu',i\omega)=
\begin{cases}
F_{a}(i\nu,i\nu',i\omega) & \text{if } |\nu-\omega/2||\nu'-\omega/2|(|\omega|+1) > f\beta, \\
F_m(i\nu,i\nu',i\omega) & \text{otherwise},
\end{cases}
\end{align}
where $\nu$ and $\nu'$ are the fermionic Matsubara frequencies and $\omega$ is the bosonic Matsubara frequency. This is motivated by Ref.~\cite{kaufmann_continuous-time_2017}. However, we use a modified criterion which we find better defines the boundary of the asymptotic region in this problem. We take $f\approx2$ here, which provides a good balance between minimizing noise in the vertex and accurately demarcating the asymptotic region, particularly in the spin channel. This is found qualitatively by examining the full vertex for discontinuities and noise near the boundary, and quantitatively by checking for convergence with $f$. We compute this vertex for fermionic frequencies up to 1500$t$, which we find is approximately the energy required to completely converge the static susceptibilities.

With this full vertex, we compute the local susceptibilities, $\chi=\chi_0+\chi_0F\chi_0$, where $\chi_0$ is the bubble diagram. Note that this local susceptibility differs at high frequencies from the measured local susceptibilities. Then, the Bethe--Salpeter equation is used to compute the lattice susceptibilities in the spin and charge channels.

We spend approximately 5000 CPU hours on the vertex measurements used in the phase diagram, which provides approximately $1.5\cdot10^{12}$ steps in the Monte Carlo algorithm from which to accumulate the estimators. For the dynamic and static response functions, we spend an order of magnitude more computer time. For the phase diagram, we identify a rough span of values in which the divergence occurs and then narrow the span to $\Delta n=0.025$, as represented by the error bars in Fig.~1 of the main text.

\bibliographystyle{apsrev}
\bibliography{ref}